\documentclass[epj,nopacs,fleqn]{svjour}
\pdfoutput=1
\usepackage{graphicx,amsmath,amssymb,slashed,cite,verbatim,multirow,url}
\usepackage{lineno}
\modulolinenumbers[5]
\usepackage{color}
\usepackage[utf8]{inputenc}
\newcommand{\nn}{\nonumber}

\newcommand{\MET}{\ensuremath{\slashed{E}_T}}

\newcommand{\ie} {{\it i.e.}}
\newcommand{\eg} {{\it e.g.}}

\newcommand {\beq} {\begin{equation}}
\newcommand {\eeq} {\end{equation}}
\newcommand {\bea} {\begin{eqnarray}}
\newcommand {\eea} {\end{eqnarray}}
\newcommand{\GeV}{{\rm\ GeV}}

\newcommand{\TeV}{{\rm\ TeV}}

\newcommand{\fbi}{{\rm\ fb}$^{-1}$}

\newcommand{\change}{\color{black}} 

\begin{document}

\title{Simplified dark matter models with a spin-2 mediator\\ 
       at the LHC} 

\author{
 Sabine Kraml\inst{1},
 Ursula Laa\inst{1,2},
 Kentarou Mawatari\inst{1,3},
 Kimiko Yamashita\inst{4}
}

\institute{ 
 Laboratoire de Physique Subatomique et de Cosmologie, 
 Universit\'e Grenoble-Alpes, CNRS/IN2P3, \\ 53 Avenue des Martyrs, F-38026 Grenoble, France
 \and
 LAPTh, Universit\'e Savoie Mont Blanc, CNRS, B.P.110 Annecy-le-Vieux, F-74941 Annecy Cedex, France
 \and
 Theoretische Natuurkunde and IIHE/ELEM, Vrije Universiteit Brussel,
 and International Solvay Institutes,\\
 Pleinlaan 2, B-1050 Brussels, Belgium
 \and
 Department of Physics, Graduate School of Humanities and Sciences,
 and Program for Leading Graduate Schools,\\  
          Ochanomizu University, Tokyo 112-8610, Japan
}

\abstract{
We consider simplified dark matter models where a dark matter candidate couples to the standard model (SM) particles via an $s$-channel spin-2 mediator, and study constraints on the model parameter space from {\change the current LHC data}.
{\change Our focus lies on} the complementarity among different searches, in particular monojet and multijet plus missing energy searches and resonance searches. 
For universal couplings of the mediator to SM particles, 
{\change missing-energy searches can give stronger constraints than $WW$, $ZZ$, dijet, dihiggs, $t\bar t$, $b\bar b$ resonance searches in the low-mass region and/or when the coupling of the mediator to dark matter is much larger than its couplings to SM particles. The strongest constraints however come from diphoton and dilepton resonance searches. Only if these modes are suppressed, missing-energy searches can be competitive in constraining dark matter models with a spin-2 mediator.}
}

\date{}

\titlerunning{Simplified dark matter models with a spin-2 mediator at the LHC}
   
\authorrunning{S.~Kraml et al.}

\maketitle

\vspace*{-11.5cm}
\noindent 
Preprint: OCHA-PP-345
\vspace*{10cm}


\section{Introduction}

Convincing astrophysical and cosmological observations for the existence of dark matter (DM) provide us one of the strong motivations to consider physics beyond the standard model (SM). The search for DM is thus one of the main pillars of the LHC physics program. 

As the nature of DM is known so little, a so-called simplified-model approach~\cite{Alves:2011wf} has been widely adopted, 
and concrete simplified DM models have recently been proposed by the LHC DM working group to conduct the systematic DM searches at the LHC Run-II~\cite{Abercrombie:2015wmb}.
Following the proposal, the Run-I data as well as 
the early Run-II data have already been analysed to constrain simplified DM models with $s$-channel spin-1 and spin-0 mediators, see \eg~\cite{Khachatryan:2016mdm,Aaboud:2016uro,Aaboud:2016tnv,Aaboud:2016qgg,Aaboud:2016obm,ATLAS-CONF-2016-086,Sirunyan:2016iap,CMS-PAS-EXO-16-037,CMS-PAS-EXO-16-038,CMS-PAS-EXO-16-039}.  
On the other hand, the model with a spin-2 mediator~\cite{Lee:2013bua,Lee:2014caa} has not been fully explored for the LHC yet---it is one of the next-generation simplified DM models~\cite{Bauer:2016gys}.

In this article, we consider simplified DM models where a DM candidate couples to the SM particles via an $s$-channel spin-2 mediator, and study constraints on the model parameter space from searches in final states with and without missing energy in the current 
LHC data. 
This work follows the {\sc DMsimp} framework~\cite{Mattelaer:2015haa,Backovic:2015soa,Neubert:2015fka}, which provides the DM model files for event generators such as {\sc MadGraph5\_aMC@NLO}~\cite{Alwall:2014hca} as well as for DM tools such as {\sc micrOMEGAs}~\cite{Belanger:2006is,Belanger:2008sj,Barducci:2016pcb} and {\sc MadDM}~\cite{Backovic:2013dpa,Backovic:2015cra}. The same framework was used previously to study the cases of  $s$-channel spin-1 and spin-0 mediators. 

We note that, to keep the analysis of the LHC constraints fully general, we do not impose any astrophysical constraints 
like relic density or (in)direct detection limits on the DM candidate, as these partly depend on astrophysical assumptions. 
Moreover, in a full model, the DM may couple to other new particles that are irrelevant for the collider phenomenology 
discussed here. We refer readers to~\cite{Lee:2013bua,Lee:2014caa} for the astrophysical constraints, and to \cite{Garcia-Cely:2016pse} for a discussion of spectral features in indirect detection. 

The article is organised as follows. 
The simplified model is presented in Section~\ref{sec:model}, 
and the production and decays of the spin-2 mediator in Section~\ref{sec:pheno}. 
The re-interpretation of the LHC results is discussed in Section~\ref{sec:results}. 
Section~\ref{sec:summary} contains a summary and conclusions.
Supplemental material for recasting is provided in the Appendix.

\section{Model}
\label{sec:model}

Gravity-mediated DM was proposed in~\cite{Lee:2013bua,Lee:2014caa}, where the dark sector communicates with the SM sector through a new spin-0 particle (radion) and spin-2 particles (Kaluza--Klein (KK) gravitons) in warped extra-dimension models as well as in the dual composite picture.

In this work, following the approach of simplified DM models, we consider DM particles which interact with the SM particles via an $s$-channel spin-2 mediator. 
The interaction Lagrangian of a spin-2 mediator ($Y_2$) with DM ($X$) is given by~\cite{Lee:2013bua}
\begin{align}\label{y-x}
 {\cal L}_{X}^{Y_2} = -\frac{1}{\Lambda} g^{T}_{X}\,T^X_{\mu\nu} Y_2^{\mu\nu}\,,
\end{align}
where $\Lambda$ is the scale parameter of the theory, $g^T_X$ is the coupling parameter, and $T_{\mu\nu}^X$ is the energy--momentum tensor of a DM field.
Here, we consider three types of DM independently; a real scalar ($X_R$), a Dirac fermion ($X_D$), and a vector ($X_V$).
The interaction with SM particles is obtained by 
\begin{align}\label{y-sm}
 {\cal L}_{\rm SM}^{Y_2} = -\frac{1}{\Lambda} \sum_i g^{T}_{i}\, T^i_{\mu\nu} Y_2^{\mu\nu}\,,
\end{align}
where $i$ denotes each SM field, \ie\ the Higgs doublet ($H$), quarks ($q$), leptons ($\ell$),
 and $SU(3)_C$, $SU(2)_L$ and $U(1)_Y$ gauge bosons ($g,W,B$).
Following~\cite{Ellis:2012jv,Englert:2012xt} we introduce the
phenomenological coupling parameters
\begin{align}
 g^T_i=\{g^T_H,\,g^T_q,\,g^T_\ell,\,g^T_g,\,g^T_W,\,g^T_B\}
\label{smcouplings}
\end{align}
without assuming any UV model.%
\footnote{One may also assign independent coupling parameters for each flavour, especially for heavy flavours~\cite{Das:2016pbk}.}
The energy--momentum tensors of the DM are 
\begin{align}\label{em-x}
 T_{\mu\nu}^{X_R} &= -\frac{1}{2}g_{\mu\nu}(
  \partial_{\rho}X_R\partial^{\rho}X_R - m^2_{X}X^2_R) \nn\\
 &\quad +\partial_{\mu}X_R\partial_{\nu}X_R\,,\\ 
 T_{\mu\nu}^{X_D} &= -g_{\mu\nu}(
  \overline{X}_Di\gamma_{\rho}\partial^{\rho}X_D - m_{X}\overline{X}_DX_D) \nn\\
 &\quad +\frac{1}{2}g_{\mu\nu}\partial_{\rho}(\overline{X}_Di\gamma^{\rho}X_D) \nn\\
 &\quad +\frac{1}{2}
  \overline{X}_D i(\gamma_{\mu}\partial_{\nu}+\gamma_{\nu}\partial_{\mu}) X_D \nn\\
  &\quad-\frac{1}{4}\partial_{\mu}(\overline{X}_Di\gamma_{\nu}X_D)
  -\frac{1}{4}\partial_{\nu}(\overline{X}_Di\gamma_{\mu}X_D)\,,\\
 T_{\mu\nu}^{X_V} &= -g_{\mu\nu}(-\frac{1}{4}F_{\rho\sigma} F^{\rho\sigma}
  + \frac{m_X^2}{2} X^{}_{V\rho}X_{V}^{\rho}) \nn\\
 &\quad +F_{\mu\rho}F^{\rho}_{\nu} +m^2_{X}X^{}_{V\mu}X^{}_{V\nu}\,,
\end{align}
where $F_{\mu\nu}$ is the field strength tensor.
Those of the SM fields are similar; see \eg~\cite{Das:2016pbk} for the explicit formulae.  

Complying with the simplified-model idea, it is instructive to 
consider universal couplings between the spin-2 mediator and the SM particles: 
\begin{align}
 g_{\rm SM}\equiv g^T_H=g^T_q=g^T_\ell=g^T_g=g^T_W=g^T_B\,.
\end{align}
With this simplification, the model has only four independent parameters,
two masses and two couplings: 
\begin{align}\label{param}
 \{m_X,\,m_Y,\,g_X/\Lambda,\,g_{\rm SM}/\Lambda\}\,,  
\end{align}
where we dropped the superscript $T$ for simplicity. 
Such a universal coupling to SM particles is realised, \eg, in the original Randall--Sundrum (RS) model of localised gravity~\cite{Randall:1999ee}.
The parameters are related as
\begin{align}
 m_Y/\Lambda=x_1\, k/\overline M_{\rm Pl}\,,
\end{align}
where $x_1=3.83$ is the first root of the Bessel function of the first kind, $k$ is the curvature of the warped extra dimension, and $\overline M_{\rm Pl}=2.4\times 10^{18}\GeV$ is the reduced four-dimensional Planck scale.
On the other hand, in the so-called bulk RS model~\cite{Agashe:2007zd,Fitzpatrick:2007qr}, where the SM particles also propagate in the extra dimension, $g^T_i$ can take different values depending on the setup.

\begin{figure}\center 
 \includegraphics[width=.88\columnwidth]{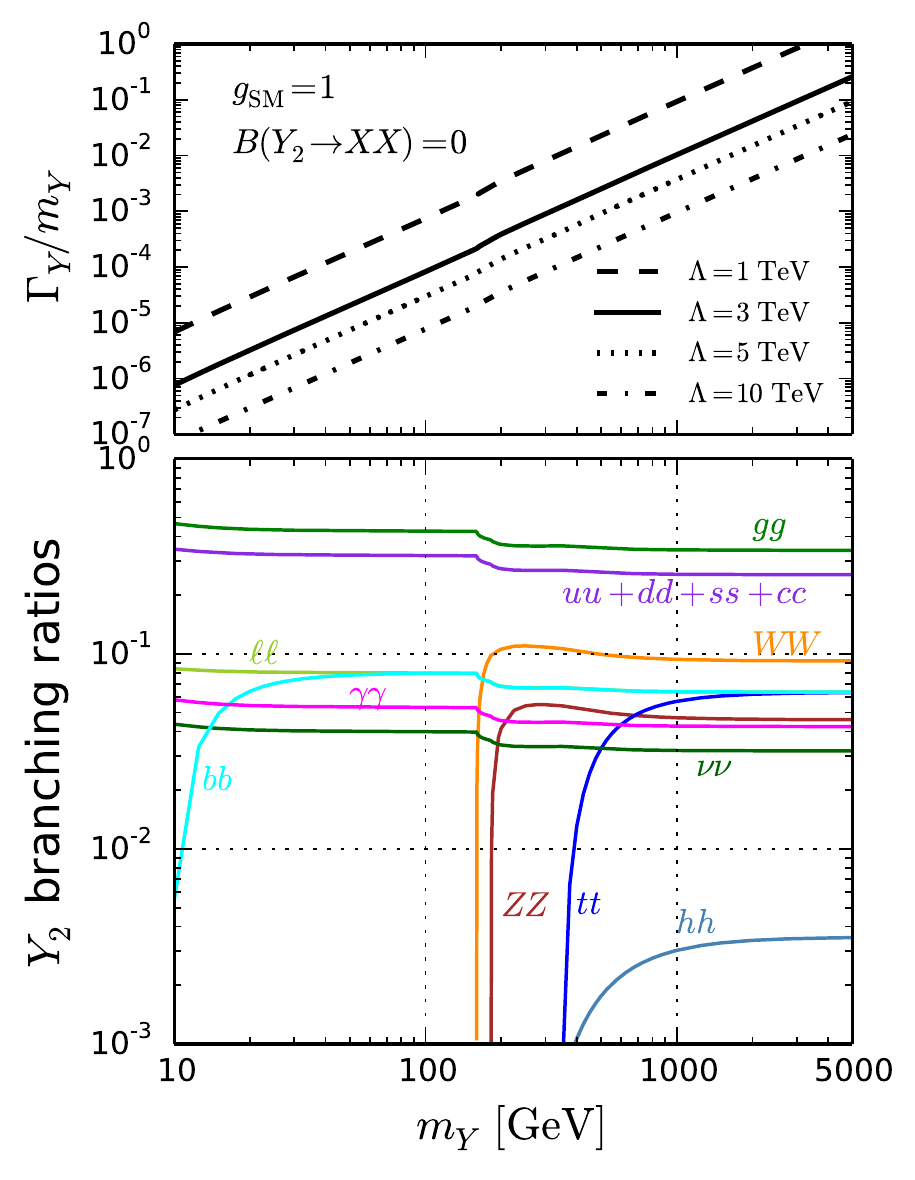}
\caption{Ratio of the mediator total width to its mass, $\Gamma_Y/m_Y$, (upper panel) and mediator branching ratios (lower panel) as a function of the mediator mass $m_Y$ for $g_{\rm SM}=1$, where we assume a negligible branching ratio to the dark sector.}
\label{fig:width_sm}
\end{figure}

\begin{table}\center
\begin{tabular}{r|rrrrrrrr}
\hline 
 $m_Y$ & \multicolumn{8}{c}{branching ratios $[\%]$} \\
 $[{\rm GeV}]$ & $jj$ & $WW$ & $tt$ & $ZZ$ & $\gamma\gamma$ & $\nu\nu$ & $ee$ & $hh$ \\
\hline
100   & 86.5   & 0 	      & 0     &     0   & 5.3  & 4.0          & 2.7     & 0 \\
500   & 79.1   & 9.9       & 3.3  & 5.0    & 4.4   & 3.3          & 2.2     & 0.2  \\
1000 & 78.5   & 9.4       & 5.7  & 4.7    & 4.3    & 3.2          & 2.1     & 0.3 \\
\hline
\end{tabular}
\caption{Branching ratios of the spin-2 mediator for $g_{\rm SM}=1$ and $B(Y_2\to XX)=0$; 
$jj$ includes gluons and five flavours of quarks, and $\nu\nu$ includes three flavours of neutrinos.}
\label{tab:width_sm}
\end{table}

In~\cite{Das:2016pbk}, 
the SM sector of the above model was implemented in {\sc FeynRules/NloCT}~\cite{Alloul:2013bka,Degrande:2014vpa} (based
on~\cite{Hagiwara:2008jb,deAquino:2011ix,Artoisenet:2013puc}),
and the $Y_2$ production and decay rates at next-to-leading order (NLO) QCD accuracy were presented.
In this work, we include the three DM species ($X_R$, $X_D$, $X_V$) with the corresponding interactions, and add the model into the {\sc DMsimp} framework~\cite{FR-DMsimp:Online} as the simplified DM model with a spin-2 mediator. 

\begin{figure*}\center 
 \includegraphics[width=0.33\textwidth]{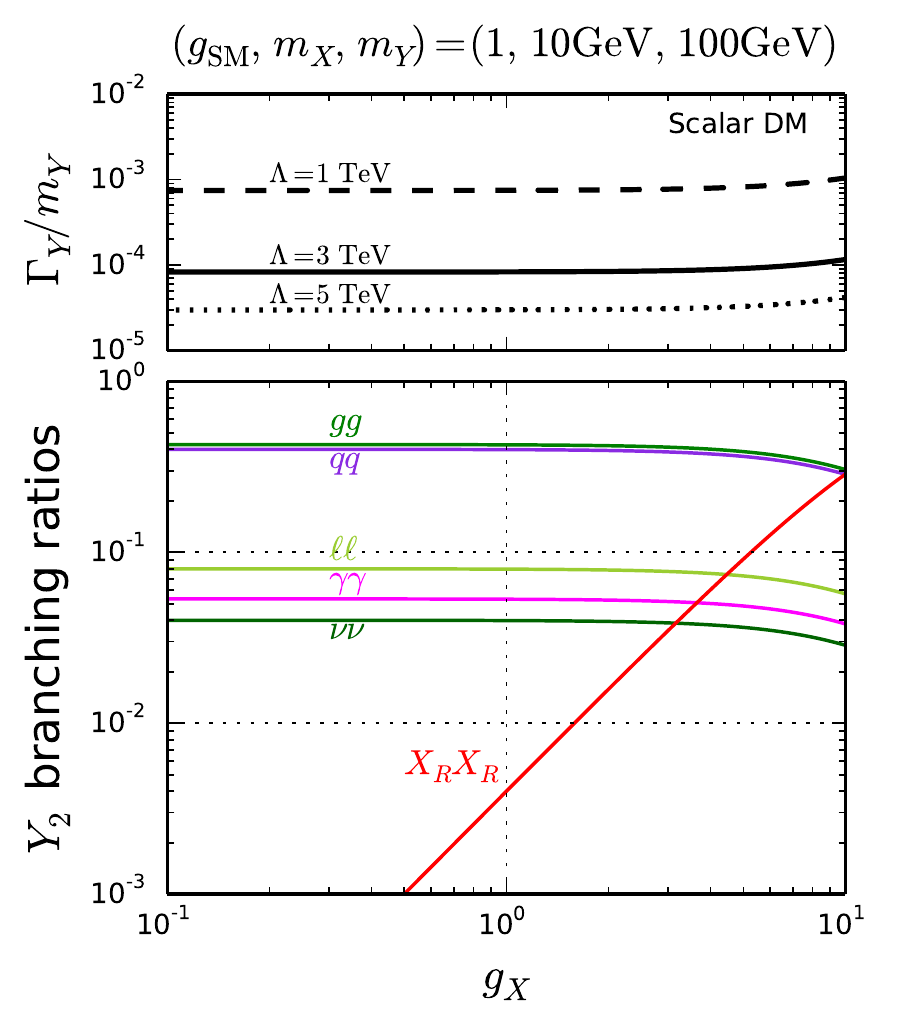}%
 \includegraphics[width=0.33\textwidth]{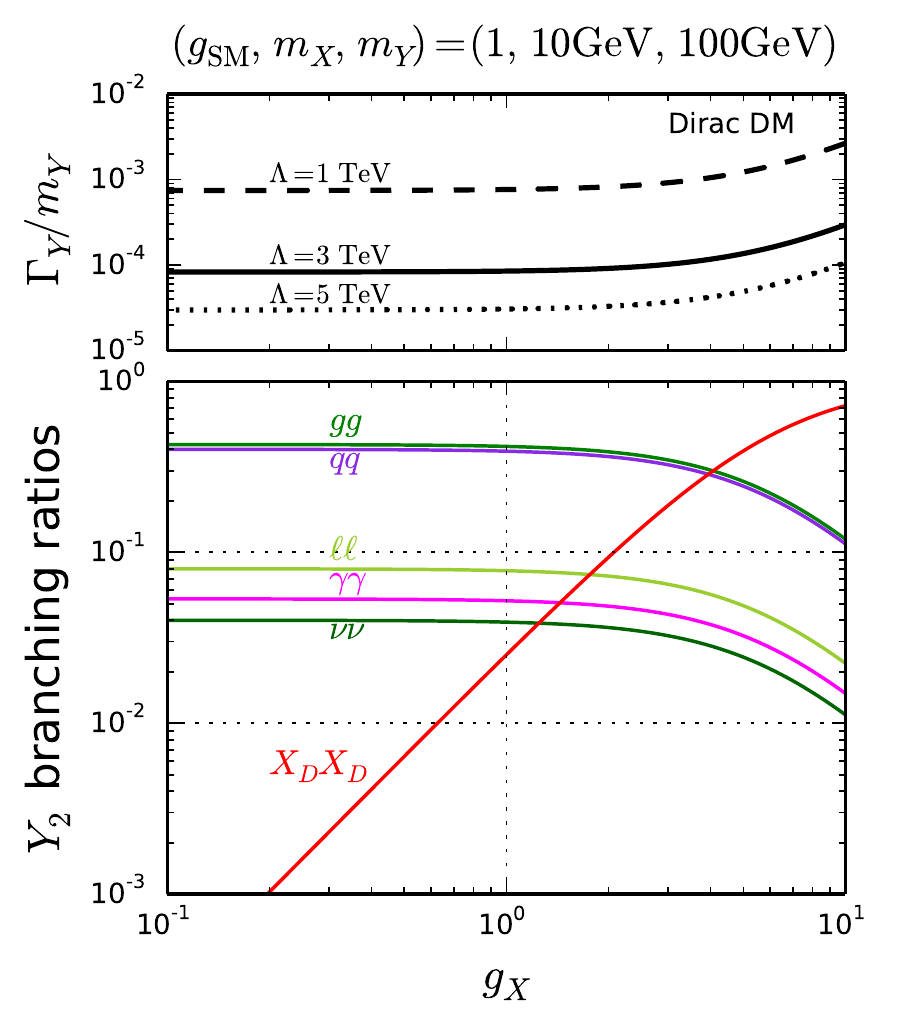}%
 \includegraphics[width=0.33\textwidth]{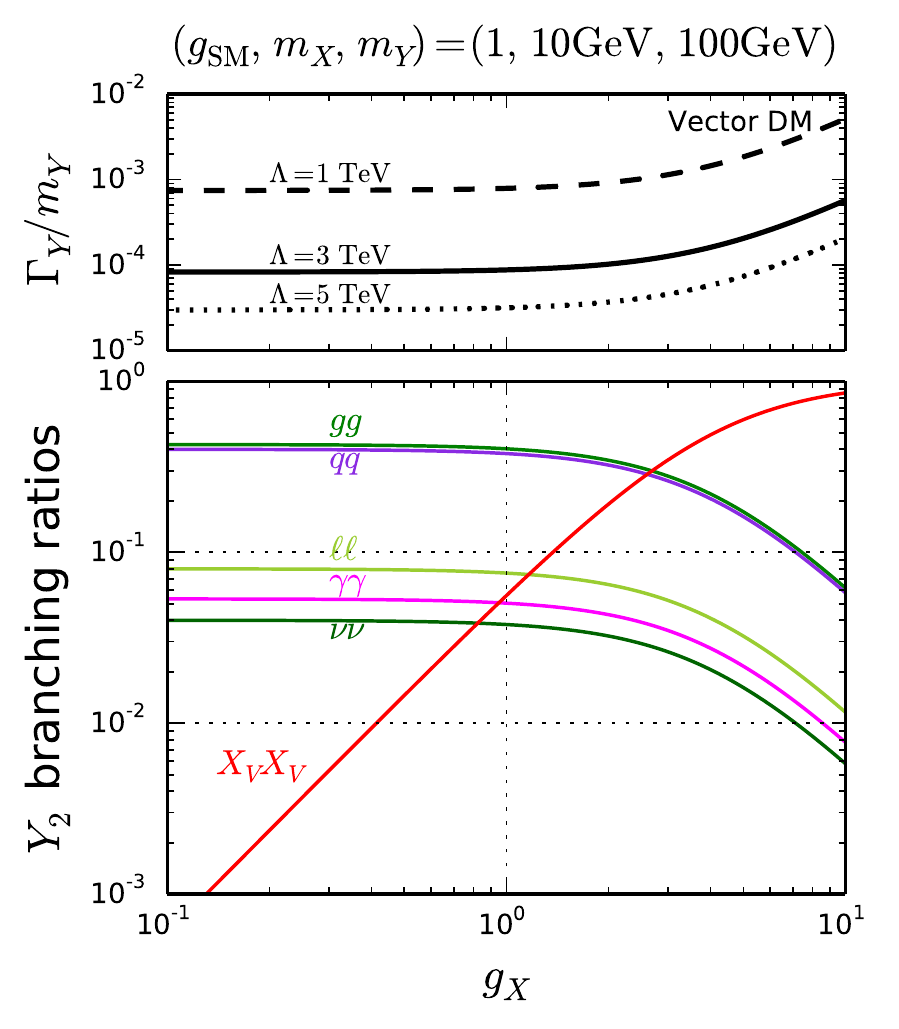}
 \includegraphics[width=0.33\textwidth]{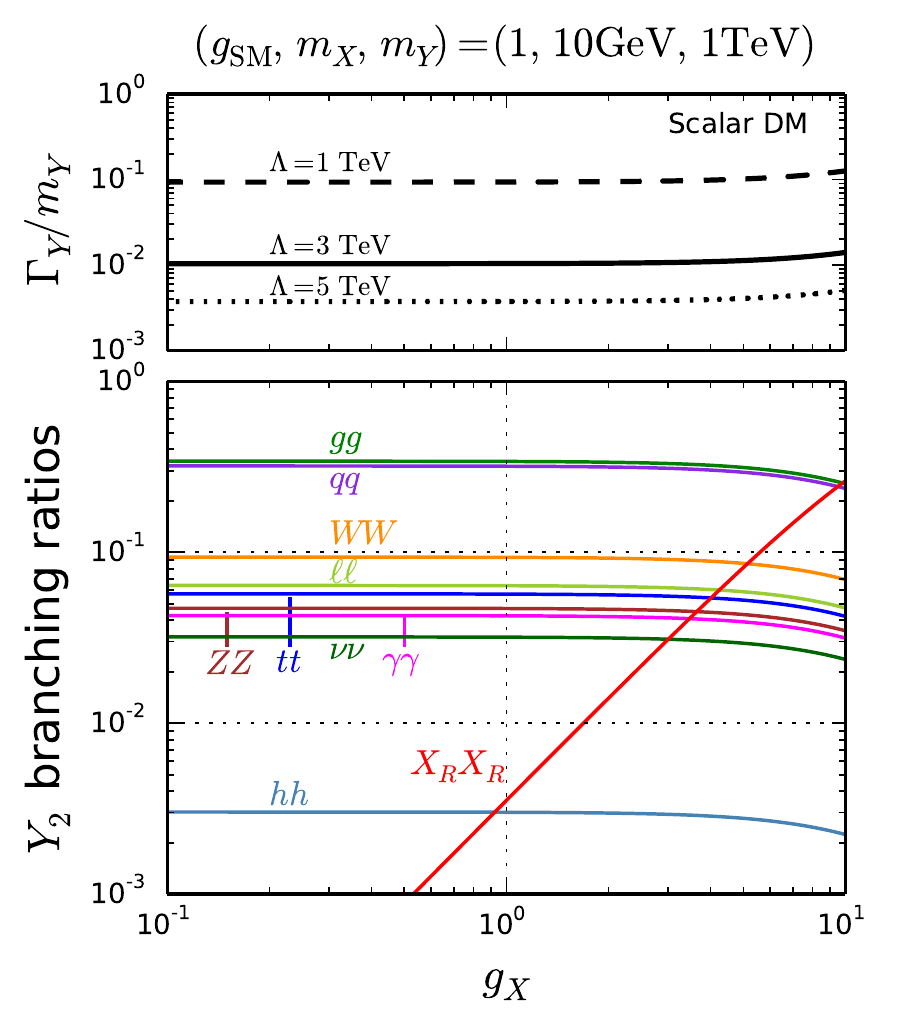}%
 \includegraphics[width=0.33\textwidth]{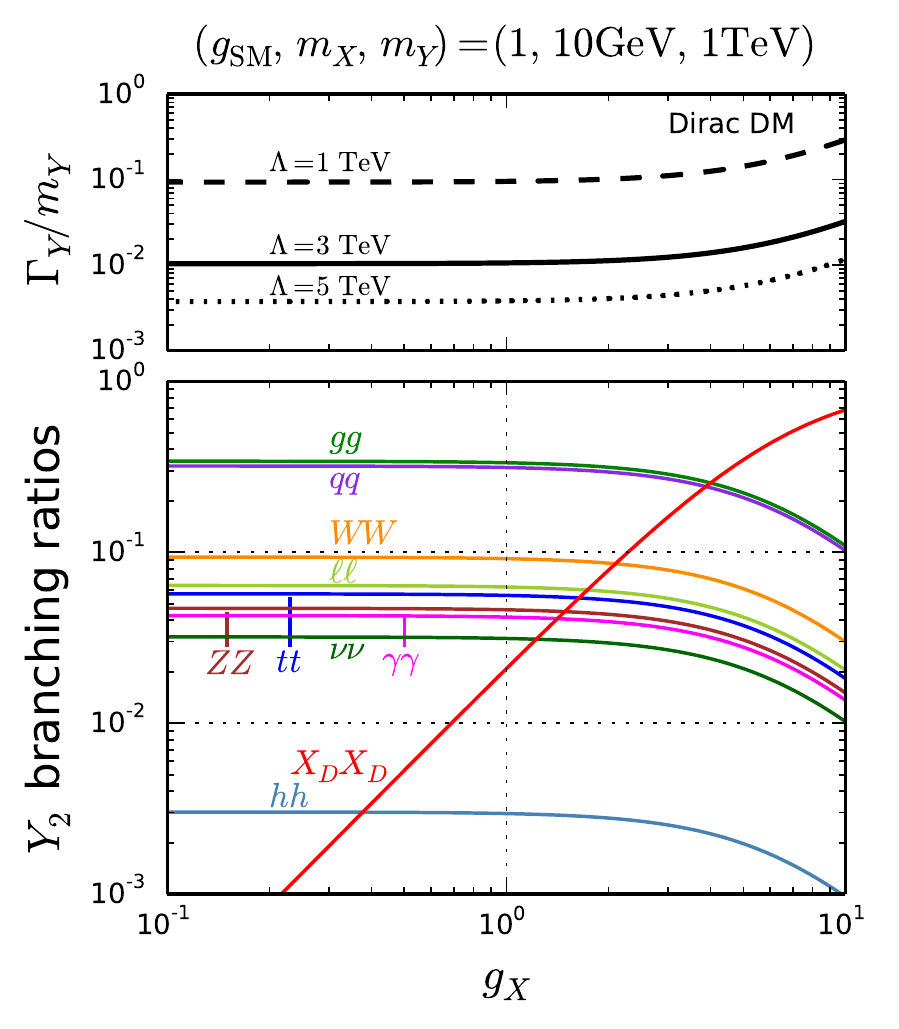}%
 \includegraphics[width=0.33\textwidth]{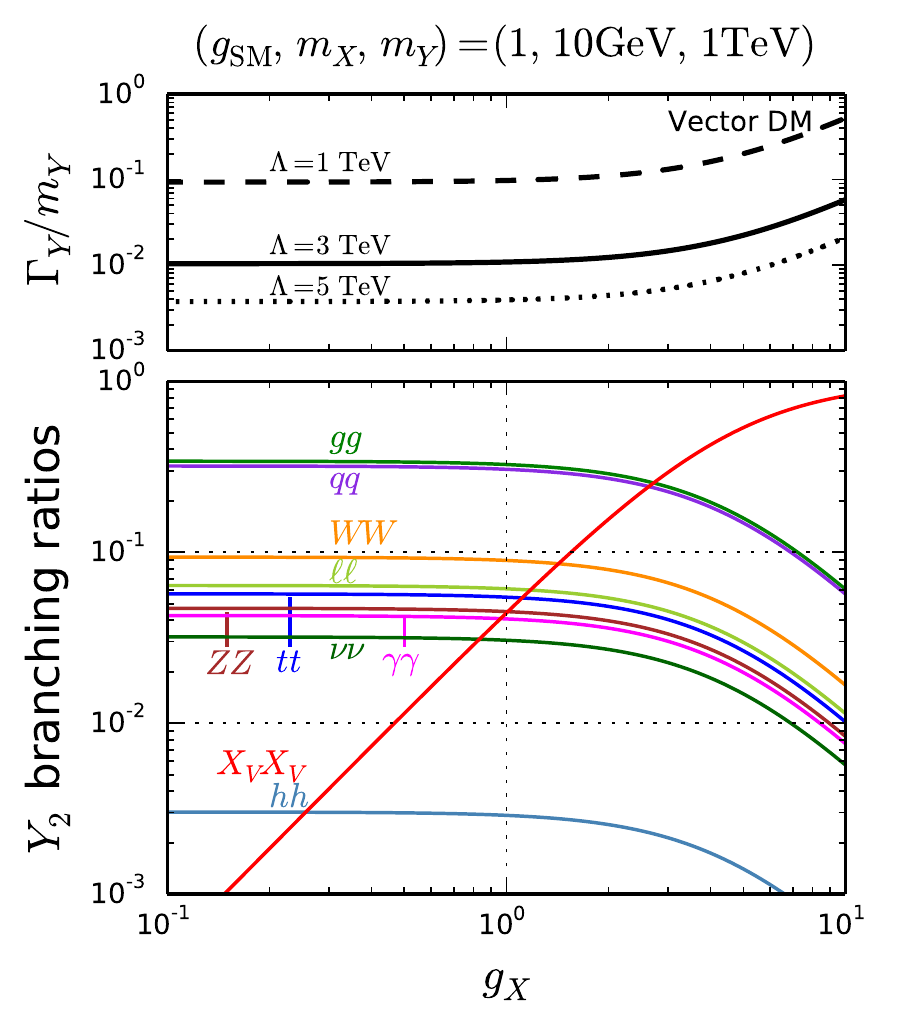} 
\caption{Ratio of the mediator total width to its mass and mediator branching ratios as a function of the DM coupling $g_X$, for mediator masses of 100~GeV (top row) and 1~TeV (bottom row). The left, middle and right columns are for scalar, Dirac and vector DM, respectively. We take $g_{\rm SM}=1$ and fix the DM mass to 10\GeV.}
\label{fig:width_dm}
\end{figure*}

\section{Phenomenology at the LHC}\label{sec:pheno}

\subsection{Decay of the spin-2 mediator}

Regarding LHC phenomenology, let us begin by discussing the spin-2 mediator decays.
The partial widths for the decays into a pair of spin-0 ($S=X_R,h$), spin-1/2 ($F=X_D,q,\ell$) and spin-1 ($V=X_V,g,\gamma,Z,W$) DM or SM particles are given by 
\begin{align}
 \label{eq:decayS}
 \Gamma_{S} &= \frac{g^2_S m^3_Y}{960\pi\Lambda^2}\,\beta_S^{5}\,,\\
 \label{eq:decayF}
 \Gamma_{F} &= \frac{g^2_F N_\nu N^F_C m^3_Y}{160\pi\Lambda^2}\,\beta_F^{3}\, (1+\frac{8}{3}r_F)\,,\\
 \label{eq:decayV}
 \Gamma_{V} &= \frac{g^2_V N_s N^V_C m^3_Y}{40\pi\Lambda^2}\,\beta_V\, f(r_V)\,, 
\end{align}
where $\beta_i=\sqrt{1-4r_i}$ with $r_i=m^2_i/m^2_Y$, $g_\gamma=g_B\cos^2\theta_W+g_W\sin^2\theta_W$ and 
$g_Z =g_B\sin^2\theta_W+g_W\cos^2\theta_W$ with the weak-mixing angle $\theta_W$, and
$f(r_V)= 1 + \frac{1}{12}\kappa^2_H 
        -r_V(3 -\frac{20}{3}\kappa_H -\kappa^2_H)
        +r^2_V(6 -\frac{20}{3} \kappa_H +\frac{14}{3} \kappa_H^2)$
with $\kappa_H=g_H/g_V$. 
For gluons and photons, $\kappa_H=0$ in $f(r_V)$, while $\kappa_H=1$ for vector DM.
The factors $N_\nu=1/2$ for neutrinos and $N_s=1/2$ for two identical particles, and are unity otherwise;
$N_C^{F,V}$ is the number of colours.
We note that $B(Y_2\to Z\gamma)=0$ for $g_W=g_B$ as 
the decay rate is proportional to 
$g_{Z\gamma}^2=[(g_W-g_B)\cos\theta_W\sin\theta_W]^2$. 
We see that, due to the different overall prefactors, the partial widths become larger in order of scalar, fermion, vector DM.
Moreover, the different powers (5, 3, 1) of the velocity factor $\beta_i$ indicate that the decay proceeds mainly via a D, P, and S wave for the scalar, fermion, and vector case, respectively.

Figure~\ref{fig:width_sm} shows the $Y_2$ total width scaled by the mass, $\Gamma_Y/m_Y$, and the decay branching ratios for the case that only decays into SM particles are allowed. 
{\sc MadWidth}~\cite{Alwall:2014bza} provides the partial decay 
rates numerically for each parameter point.
In Table~\ref{tab:width_sm} we provide the explicit values for a few representative mass points.
We see that, for a universal coupling $g_{\rm SM}$, decays into gluons and light quarks, leading to a dijet signature, are completely dominant ($\gtrsim 80\%$ depending on $m_Y$).
The diphoton channel has 4--5\% branching ratio;
other diboson channels ($WW$ and $ZZ$) as well as $t\bar t$ are important as well when kinematically allowed. Finally, it is important to note that decays into neutrinos have 3--4\% branching ratio, leading to missing energy signatures independent of decays to DM.%
\footnote{These decay branching ratios were already presented in~\cite{Allanach:2002gn} for the case of the RS graviton. We repeat them here for the sake of completeness. Our numbers agree with~\cite{Allanach:2002gn} apart from a factor 1/2 for decays into neutrinos.}
The width is proportional to $m_Y^3$, and  
from the upper panel in Fig.~\ref{fig:width_sm} we see that for $g_{\rm SM}/\Lambda\lesssim(3\TeV)^{-1}$,  the resonance is always very narrow
($\Gamma_Y/m_Y<1\%$) up to $m_Y\sim 1\TeV$.
Note here, that $\Lambda$ is simply a scale parameter, not a physical cut-off of the theory. 

When decays into DM are allowed, their relative importance depends on $g_X$ and the type of DM (scalar, Dirac or vector) as illustrated in Fig.~\ref{fig:width_dm}; see also Eqs.~\eqref{eq:decayS}--\eqref{eq:decayV}.
Two mass scales are considered:
$m_Y=100$~GeV and 1~TeV, with $m_X=10$~GeV and $g_{\rm SM}=1$.%
\footnote{As can be deduced from Fig.~\ref{fig:width_sm}, above the $WW$ threshold up to high masses the picture does not change much apart from the $t\bar t$ and/or $hh$ channels being open or not.}  
We see that decays into DM can be important and even dominant, but the resonance remains narrow for any choice of $\Lambda \gtrsim 3$~TeV for $m_Y\lesssim1\TeV$.
Another important observation is that for scalar DM ($X_R$), for $g_X\sim g_{\rm SM}$ the decay into $Y_2\to X_RX_R$ is practically irrelevant; one needs $g_X/g_{\rm SM}\approx 3$
for the decay into DM to exceed the one into neutrinos, and $g_X/g_{\rm SM}\approx 5$--$6$ to reach the 10\% level. 
For Dirac ($X_D$) and vector ($X_V$) DM, the decays into DM and into neutrinos are of comparable magnitude at $g_X\sim g_{\rm SM}$, both contributing to missing-energy signatures. For $g_X/g_{\rm SM}=2$, the branching ratio of $Y_2\to X_DX_D$ $(X_VX_V)$ attains about 10\% (20\%). These differences depending on the type of DM will be important later for the collider limits.

\subsection{Production of the spin-2 mediator}\label{sec:production}

\begin{figure}\center
 \includegraphics[width=1.\columnwidth]{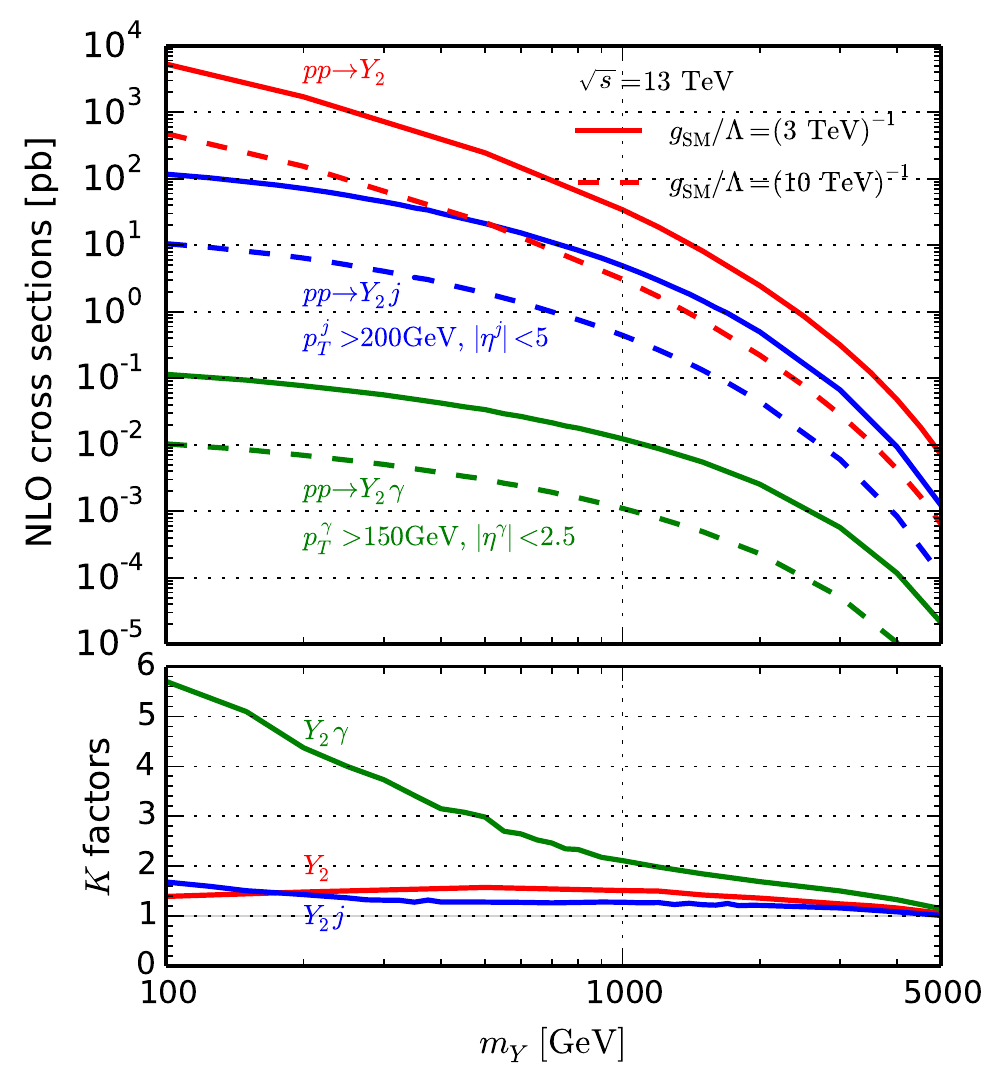}
\caption{Total cross sections at NLO accuracy for mediator productions at the 13~TeV LHC as a function of the mediator mass. 
Two choices of $g_{\rm SM}/\Lambda$ are considered: $(3\TeV)^{-1}$ shown as solid lines and $(10\TeV)^{-1}$ shown as dashed lines.
For $Y_2+{\rm jet}$ cuts of $p_T^j>200$~GeV and $|\eta^j|<5$ are imposed, and for $Y_2+{\rm photon}$ cuts of $p_T^\gamma>150$~GeV and $|\eta^\gamma|<2.5$.  
$K$ factors are also shown in the lower panel as a reference. 
 }
\label{fig:xsec-Y2}
\end{figure}

\begin{figure}\center 
 \includegraphics[width=1.\columnwidth]{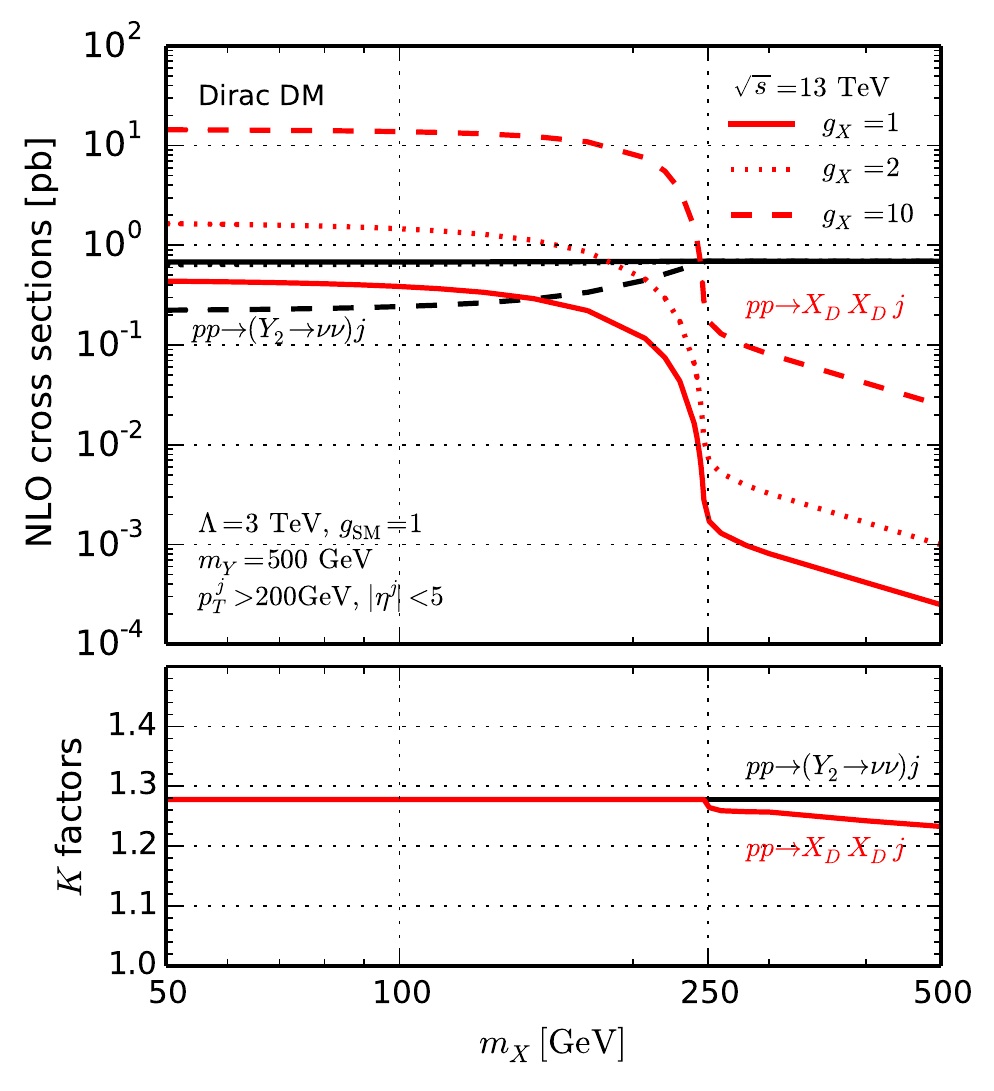}
\caption{
Total cross sections at NLO accuracy for monojet final states with $g_X=1$ (solid), 2 (dotted) and 10 (dashed) for {\change $m_Y=500$~GeV} as a function of the DM mass, where we take $\Lambda=3$~TeV and $g_{\rm SM}=1$ and impose $p_T^j>200$~GeV and $|\eta^j|<5$.
The red lines are for the (Dirac) DM channel, the black lines for the neutrino.
$K$ factors are also shown in the lower panel as a reference. 
}
\label{fig:xsec-met}
\end{figure}

Turning to the production modes, the potentially interesting channels are inclusive $Y_2$ production ($pp\to Y_2$), as well as the production with an extra hard tagging jet ($pp\to Y_2\,j$) or an electroweak boson (\eg\ $pp\to Y_2\,\gamma$). 
With the $Y_2$ decaying into SM particles, the former gives resonant peak signatures (without missing energy). 
On the other hand, the latter two give 
the typical monojet or monophoton signatures when the mediator decays invisibly.
Moreover, the latter two play a role in the low-mass resonance search in dijet events with initial-state radiation (ISR) as seen later.

The $Y_2$ production cross sections at NLO QCD accuracy
for $pp$ collisions at 13~TeV are depicted in Fig.~\ref{fig:xsec-Y2} as a function of the mediator mass.%
\footnote{See also Fig.~\ref{fig:exclusion_data_resonances}\,(bottom) for $\sigma(pp\to Y_2)$ at $\sqrt{s}=8$~TeV.} 
We employ {\sc MadGraph5\_aMC@NLO}~\cite{Alwall:2014hca} to calculate the cross sections and generate events with the LO/NLO {\sc NNPDF2.3}~\cite{Ball:2012cx}. 
The factorisation and renormalisation scales are taken at the sum of the transverse masses of the final states as a dynamical scale choice.
In our simplified model, the cross sections depend solely on $g_{\rm SM}/\Lambda$ and scale with $(g_{\rm SM}/\Lambda)^2$. 
The dashed lines showing $g_{\rm SM}/\Lambda=(10\TeV)^{-1}$ are therefore an order of magnitude below the corresponding solid lines for $g_{\rm SM}/\Lambda=(3\TeV)^{-1}$. 
Also noteworthy is the fact that $pp\to Y_2$ is mostly gluon-initiated for the low-mass case~\cite{Allanach:2002gn}; 
97\%, 83\%, and 28\% of the LO total rate for $m_Y=100\GeV$, $1\TeV$, and $5\TeV$, respectively, stem from $gg$ fusion.
Since the radiation of an initial-state photon ($Z/W$) can only occur in the quark-initiated process, $Y_2+{\rm photon}\,(Z/W)$ production is very much suppressed as compared to $Y_2+{\rm jet}$ production.
This is also the reason that the process has a huge $K$ factor especially in the low-mass region~\cite{Das:2016pbk}.%
\footnote{The $K$ factors in Fig.~\ref{fig:xsec-Y2} are slightly different from the ones reported in~\cite{Das:2016pbk} due to different PDF choices and different kinematical cuts. 
See \cite{Das:2016pbk} for details on theoretical uncertainties.} 

In the context of DM searches, the monojet signature is expected to give important constraints on the model. 
The fiducial cross sections for $pp\to Y_2j$ with $p_T^j>200$~GeV and $|\eta^j|<5$ are shown in Fig.~\ref{fig:xsec-Y2}, where one can estimate the monojet cross section by taking into account the $Y_2$ branching ratio into DM particles (and/or neutrinos) when $m_Y>2m_X$.
In Fig.~\ref{fig:xsec-met} we also plot the fiducial cross sections for $pp\to j+\MET$ as a function of the DM mass, separating the contributions from neutrinos (black lines) and DM (red lines) produced through the spin-2 mediator.
For definiteness, we take {\change $m_Y=500\GeV$}, $\Lambda=3\TeV$, $g_{\rm SM}=1$ and compare $g_X=1$, 2 and 10 for Dirac DM.
As already seen in Fig.~\ref{fig:width_dm}, their relative importance depends on $g_X$.
For $m_Y<2m_X$, a pair of DM is produced via the off-shell mediator and the cross section is strongly suppressed.
Therefore, the neutrino contribution always dominates the monojet signature for the $m_Y<2m_X$ region even if $g_X/g_{\rm SM}=10$.
For the other DM types, scalar and vector, the picture is similar, but the relative importance to the neutrino channel is different; see Fig.~\ref{fig:width_dm}. 
This is one of the  characteristic features of the spin-2 mediator DM model with universal couplings, as compared to the $s$-channel spin-1 and spin-0 models, whose mediators do not couple to charged leptons and neutrinos in the minimal setup~\cite{Abercrombie:2015wmb}.

\section{Constraints from {\change current LHC} data}\label{sec:results}

\subsection{Searches with missing energy}

The ATLAS and CMS experiments have been searching for new physics in a large variety of final states. 
As mentioned above, in the context of DM searches, the monojet signature is regarded as particularly interesting. 
In practice, at 13~TeV, the monojet analyses require one hard jet recoiling against \MET, but allow for additional jets from QCD radiation. Therefore one can expect that multijet+\MET\ searches are also relevant~\cite{Haisch:2013ata,Buchmueller:2015eea}. 

To work out the current constraints on the spin-2 mediator DM model from these searches, we consider the following early Run-II analyses: 
\begin{itemize}
 \item ATLAS monojet with $3.2$\fbi~\cite{Aaboud:2016tnv},
 \item ATLAS 2--6 jets + \MET\ with $3.2$\fbi~\cite{Aaboud:2016zdn}.
\end{itemize}
In the monojet analysis~\cite{Aaboud:2016tnv}, a simplified DM model with an $s$-channel spin-1 mediator is considered.  
Events are required to have at least one hard jet with $p_T>250\GeV$ and $|\eta|<2.4$, and a maximum of four jets with $p_T>30\GeV$ and $|\eta|<2.8$ are allowed.
Several inclusive and exclusive signal regions (SRs) are considered with increasing $\MET$ requirements from 250~GeV to 700~GeV.
The multijet+\MET\ analysis~\cite{Aaboud:2016zdn} is designed to search for squarks and gluinos in supersymmetric models, where neutralinos lead to missing energy.
Several SRs are characterised by minimum jet multiplicity from two to six; 
$\MET>200\GeV$ is required for all SRs, while different thresholds are applied on jet momenta and on the azimuthal separation between jets and $\MET$.  

To reinterpret the above analyses in the context of our spin-2 mediator simplified DM 
model, we use {\sc CheckMATE2}~\cite{Dercks:2016npn}, which is a public recasting tool providing 
confidence limits from simulated signal events and includes a number of 13~TeV analyses. 
We generate hadron-level signal samples 
by using the tree-level matrix-element plus parton-shower (ME+PS) merging procedure.
In practice, we make use of the shower-$k_T$ scheme~\cite{Alwall:2008qv}, implemented in 
{\sc MadGraph5\_aMC@NLO} \cite{Alwall:2014hca} with {\sc Pythia6} \cite{Sjostrand:2006za}, and  
generate signal events with parton multiplicity from one to two partons.
We impose $\MET>200\GeV$ and set $Q_{\rm cut}=200$~GeV for the merging separation parameter at the parton level; these values are chosen for an efficient event generation without affecting the final results. 
The event rate is normalised to the $pp\to Y_2j$ NLO cross sections shown in Fig.~\ref{fig:xsec-Y2}.
(Note, however, that NLO corrections may also affect the shapes of the kinematic distributions, as shown for the spin-1 and spin-0 cases in~\cite{Backovic:2015soa}; a detailed study of this aspect will be reported elsewhere.)

\begin{figure*}\center  
 \includegraphics[width=0.33\textwidth]{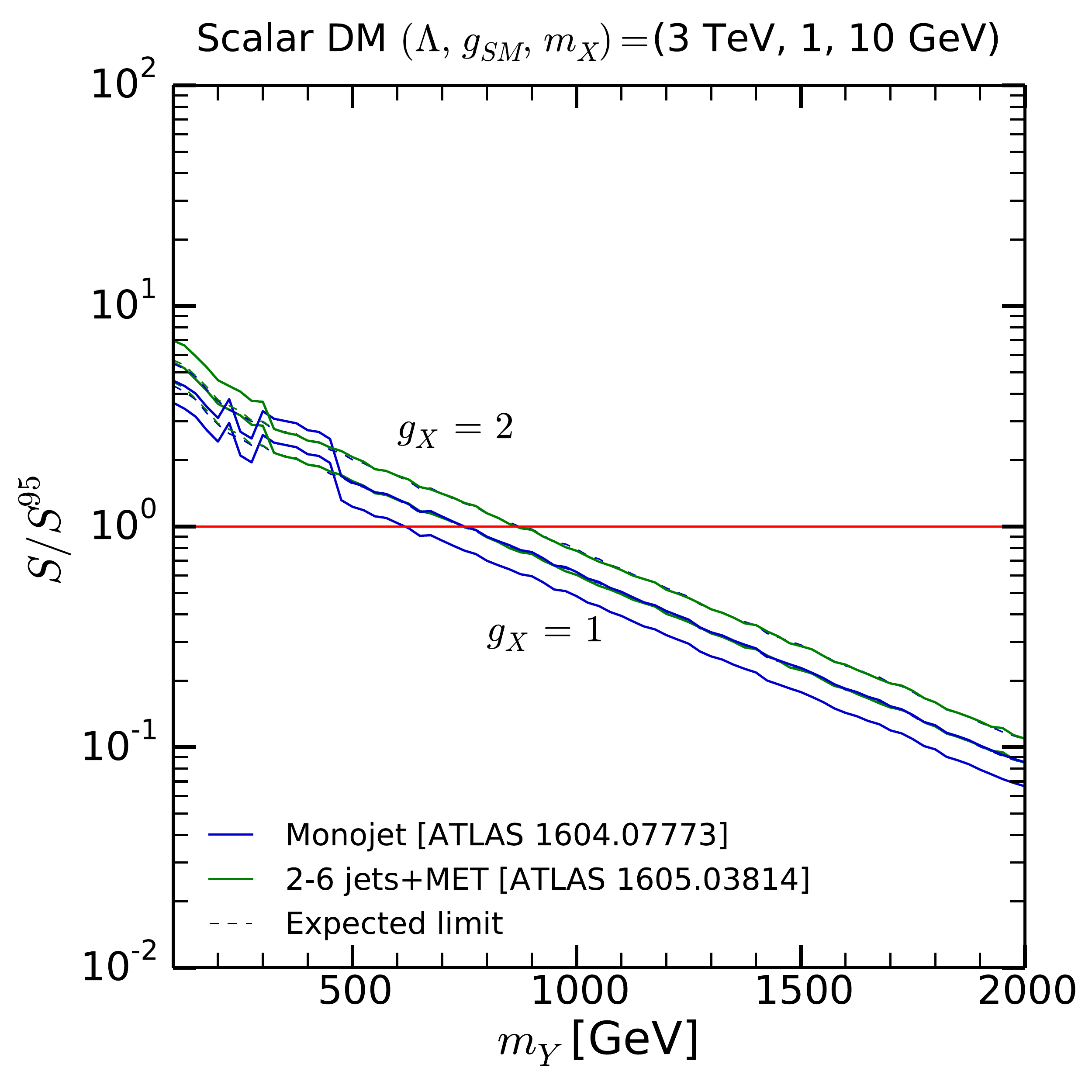}%
 \includegraphics[width=0.33\textwidth]{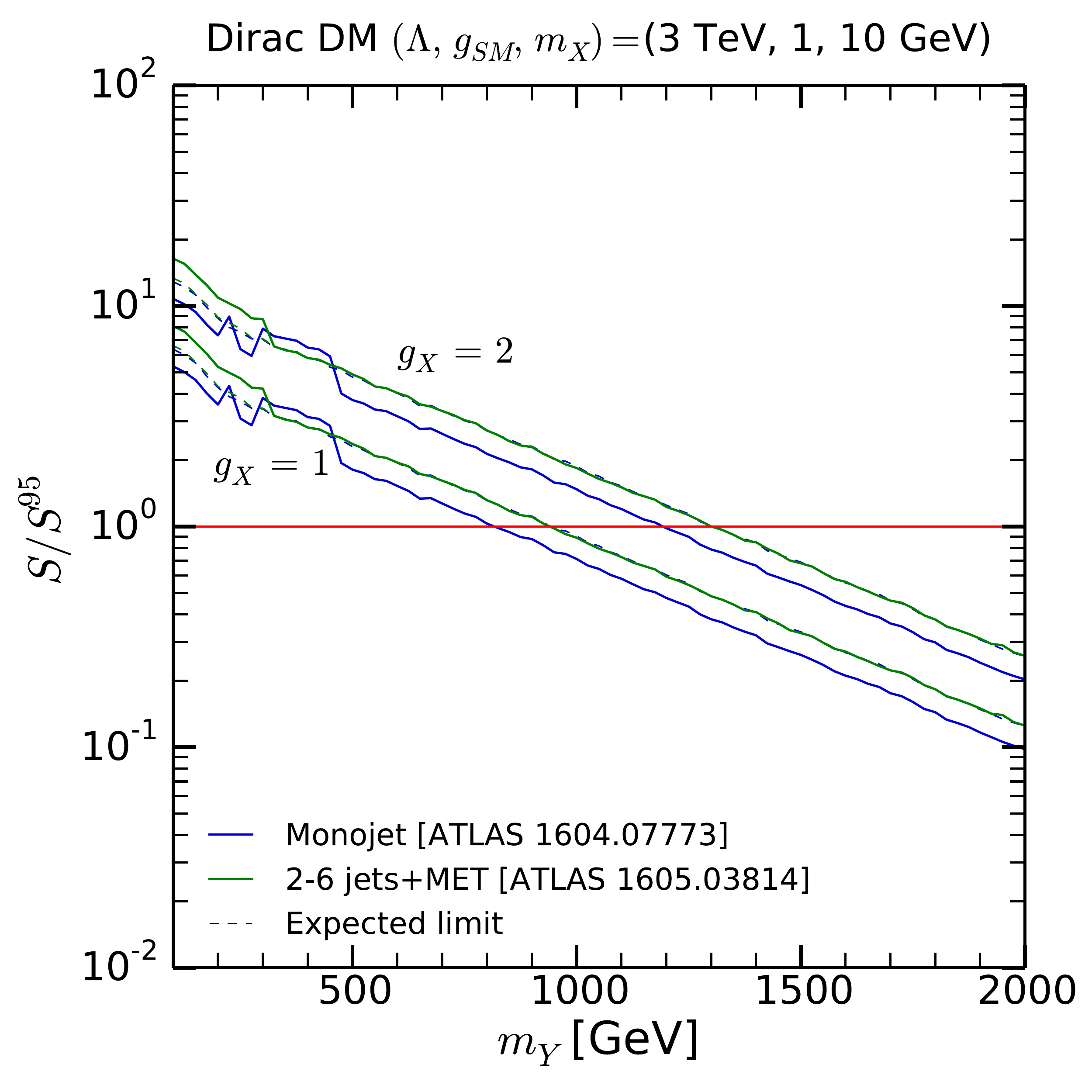}%
 \includegraphics[width=0.33\textwidth]{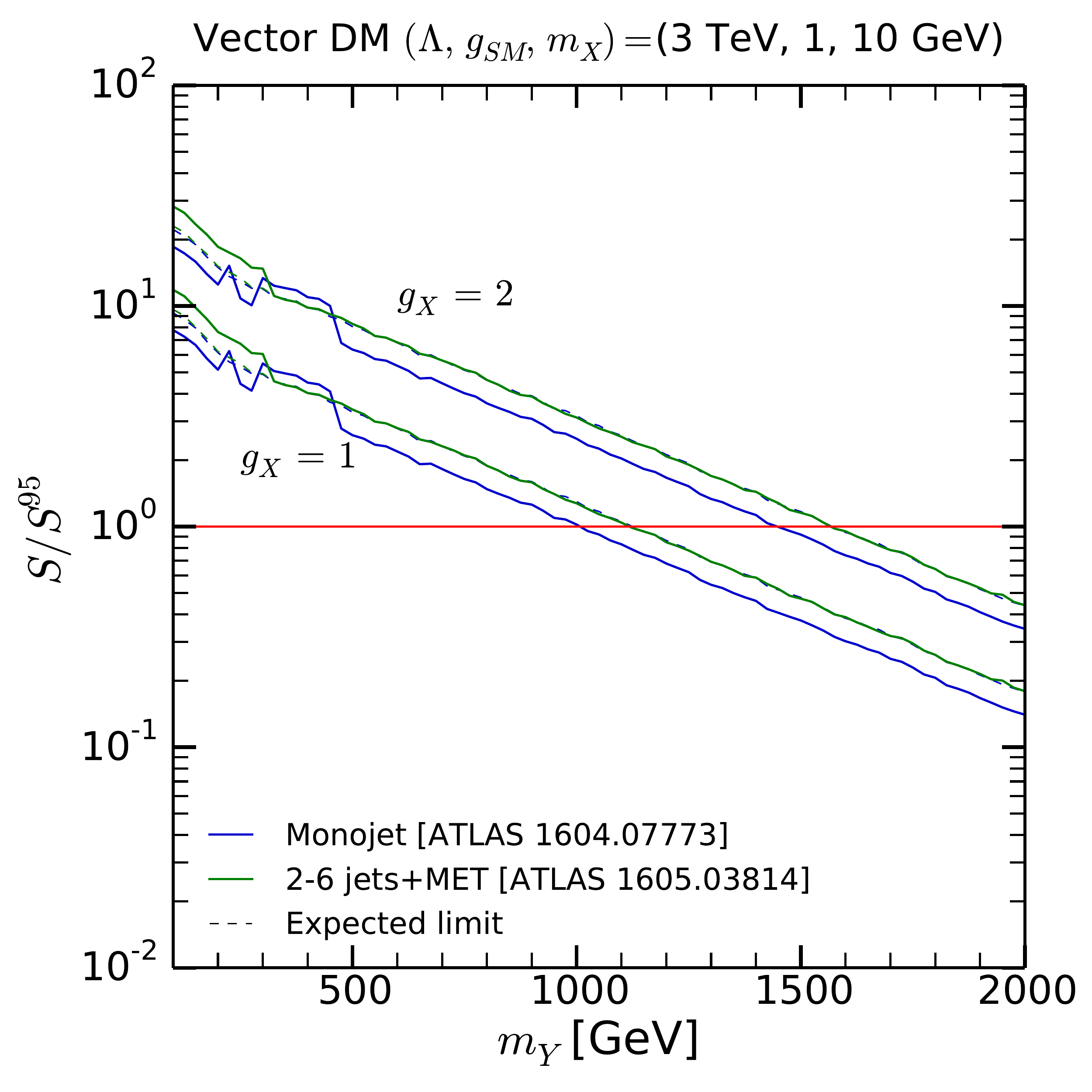}
\caption{Ratio of signal events over the number of events excluded at 95\%~CL as a function of the mediator mass, for $g_X=1$ or 2 with $\Lambda=3\TeV$, $g_{\rm SM}=1$ and $m_X=10$~GeV, where the ATLAS 13 TeV (3.2~fb$^{-1}$) monojet~\cite{Aaboud:2016tnv} and multijet+\MET~\cite{Aaboud:2016zdn} analyses are considered.
From left to right: scalar, Dirac and vector DM.
}
\label{fig:exclusion_1d_nu}
\end{figure*}

\begin{figure*}\center 
 \includegraphics[width=0.33\textwidth]{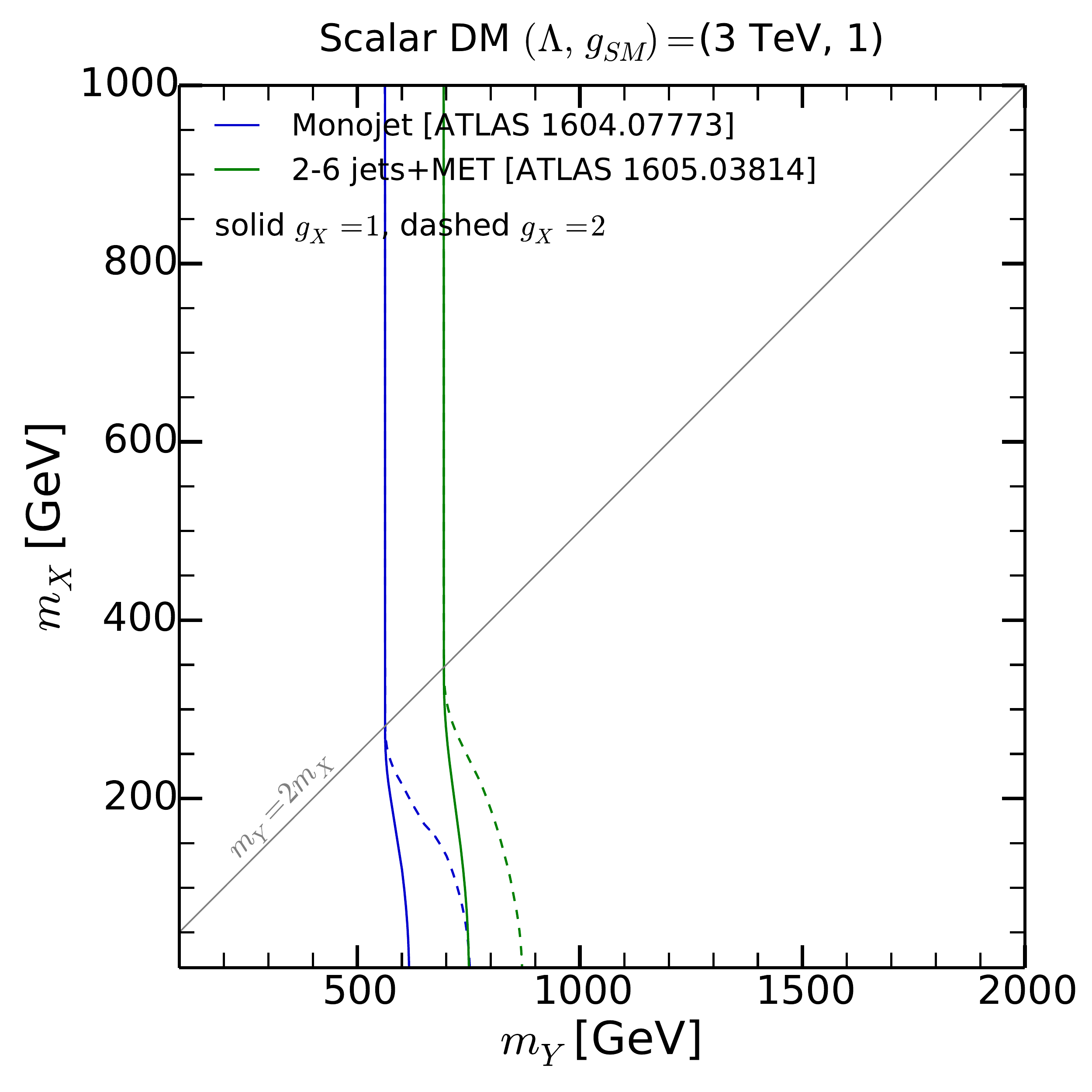}%
 \includegraphics[width=0.33\textwidth]{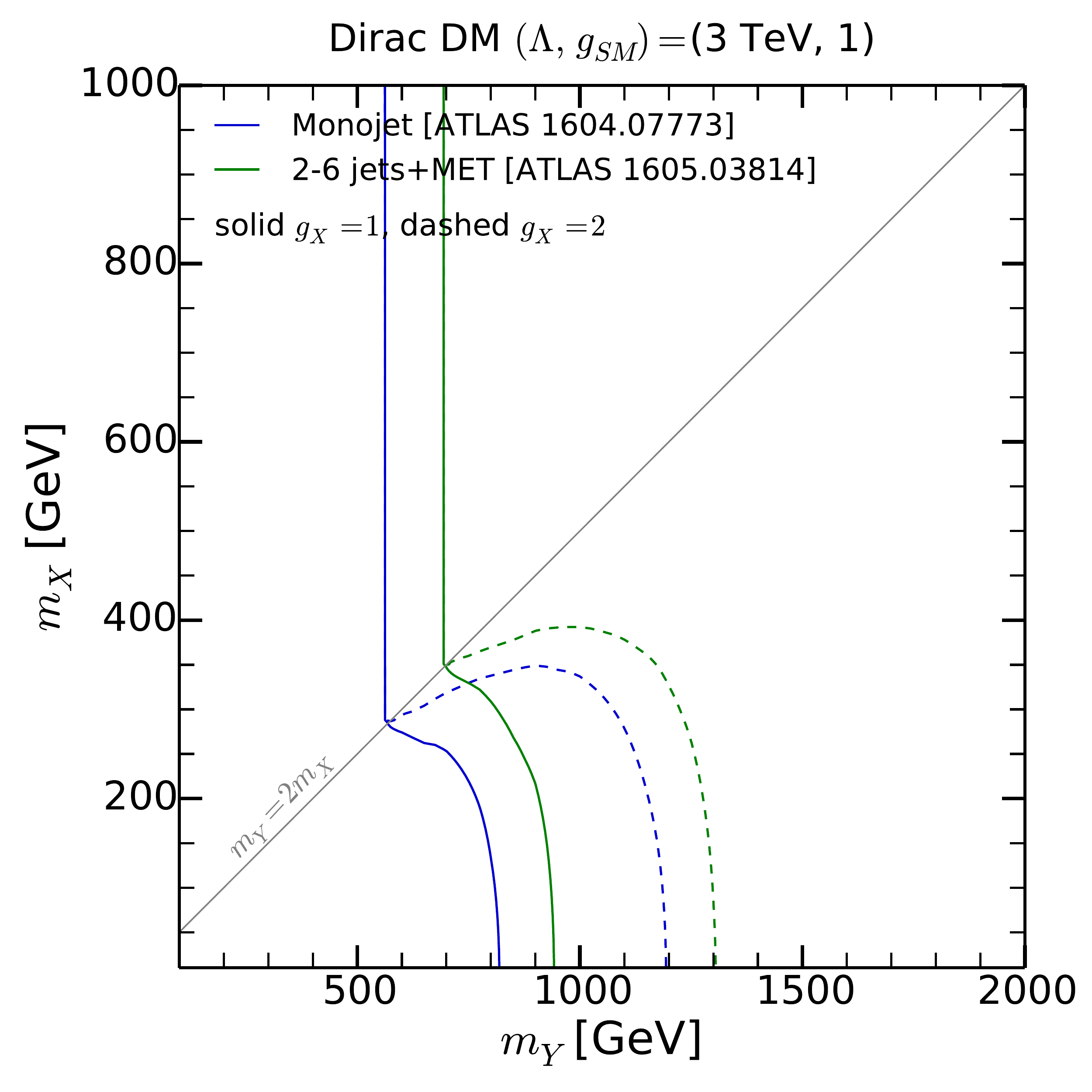}%
 \includegraphics[width=0.33\textwidth]{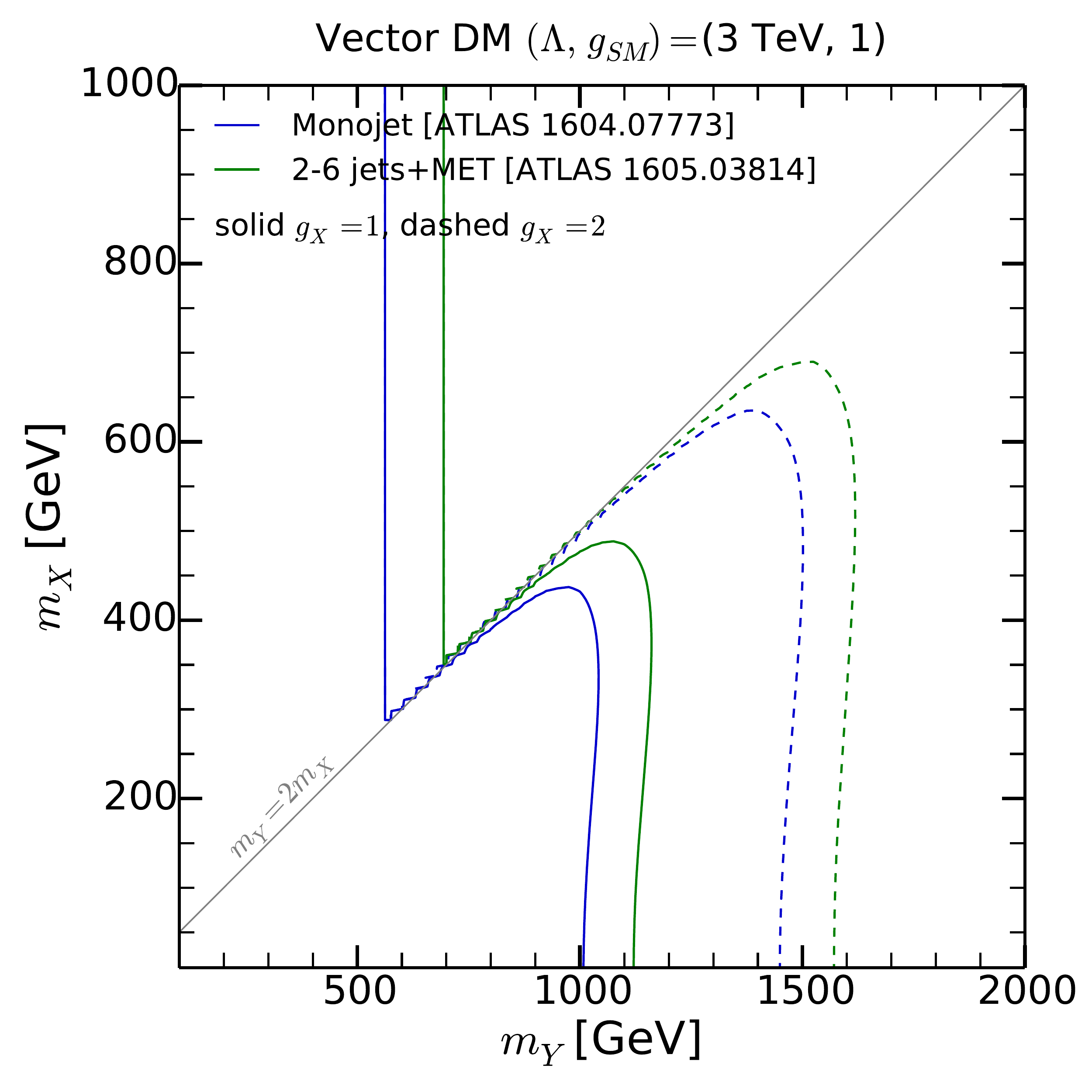}
\caption{95\% CL exclusion from  the ATLAS 13 TeV (3.2~fb$^{-1}$) monojet~\cite{Aaboud:2016tnv} and multijet+\MET~\cite{Aaboud:2016zdn} analyses in the plane of the DM vs.\ mediator masses, 
for $g_X=1$ or 2 with $\Lambda=3\TeV$ and $g_{\rm SM}=1$. 
From left to right: scalar, Dirac and vector DM.}
\label{fig:exclusion_2d_nu}
\end{figure*}

It turns out that, for an on-shell mediator of given mass, the selection efficiencies are independent of
the mass and spin of the invisible decay products. Moreover, contributions from off-shell production are negligible 
for the scenarios considered here.  The efficiencies can thus be evaluated as a function of the mediator mass only; 
see also Appendix~\ref{sec:app1}. In the following, we normalise the
number of events with NLO cross sections, shown in Fig.~\ref{fig:xsec-Y2}, and the total branching ratio into invisible final states (DM and neutrino). 
We note that for a given mediator mass the leading jet for the spin-2 mediator case is harder and more forward than that for the spin-1 case.
This is partly because the spin-2 mediator with a parton is produced not only through the $q\bar q$ and $qg$ initial states but also dominantly through the $gg$ initial state, and partly because the spin-2 mediator is also emitted from a gluon as well as from the $gggY_2$ and $q\bar qgY_2$ four-point vertices.

Figure~\ref{fig:exclusion_1d_nu} shows the ratio of signal events over the number of events excluded at 95\%~confidence level (CL), $S/S^{95}$, as a function of the mediator mass, for the three types of DM (taking $g_X=1$ or 2 with $\Lambda=3\TeV$, $g_{\rm SM}=1$ and $m_X=10$~GeV as a benchmark case). 
As expected from the discussion in the previous section, the scalar DM case is the least constrained, with the \MET\ coming dominantly from the neutrino channel;
for $g_X=1$ (2),  we find the limit $m_Y\gtrsim 600$ (750)~GeV from the monojet analysis and 
$m_Y\gtrsim 750$ (850)~GeV from the multijet+\MET\ analysis.%
\footnote{While both analyses have very similar sensitivity, \ie\ their expected limits are basically the same, the monojet results have over- and under-fluctuations in some SRs. Therefore the expected and observed limits slightly differ from each other for the monojet analysis. Overall, the multijet+\MET\ analysis tends to give the stronger limit.}
For Dirac DM the limit increases to $m_Y\gtrsim 950$ (1300)~GeV owing to the contribution from $Y_2\to X_DX_D$. 
Finally, for vector DM we have $m_Y\gtrsim 1100$ (1550)~GeV.
For the monojet analysis, the inclusive SR with the \MET\ cut of 500, 600, and 700\GeV\ (denoted as IM5, IM6, and IM7 in~\cite{Aaboud:2016tnv}) gives the limit for the low ($100\sim300\GeV$), middle ($300\sim450\GeV$), and high ($\gtrsim450\GeV$) mass region, respectively.
For the multijet+\MET\ analysis, the 2-jet loose (2jl) SR gives the limit for the mass range of $100\sim300\GeV$, while the 2-jet medium (2jm) SR does for $\gtrsim300\GeV$.
See~\cite{Aaboud:2016zdn} for the detailed selection criteria.
 
\begin{figure*}\center 
 \includegraphics[width=0.33\textwidth]{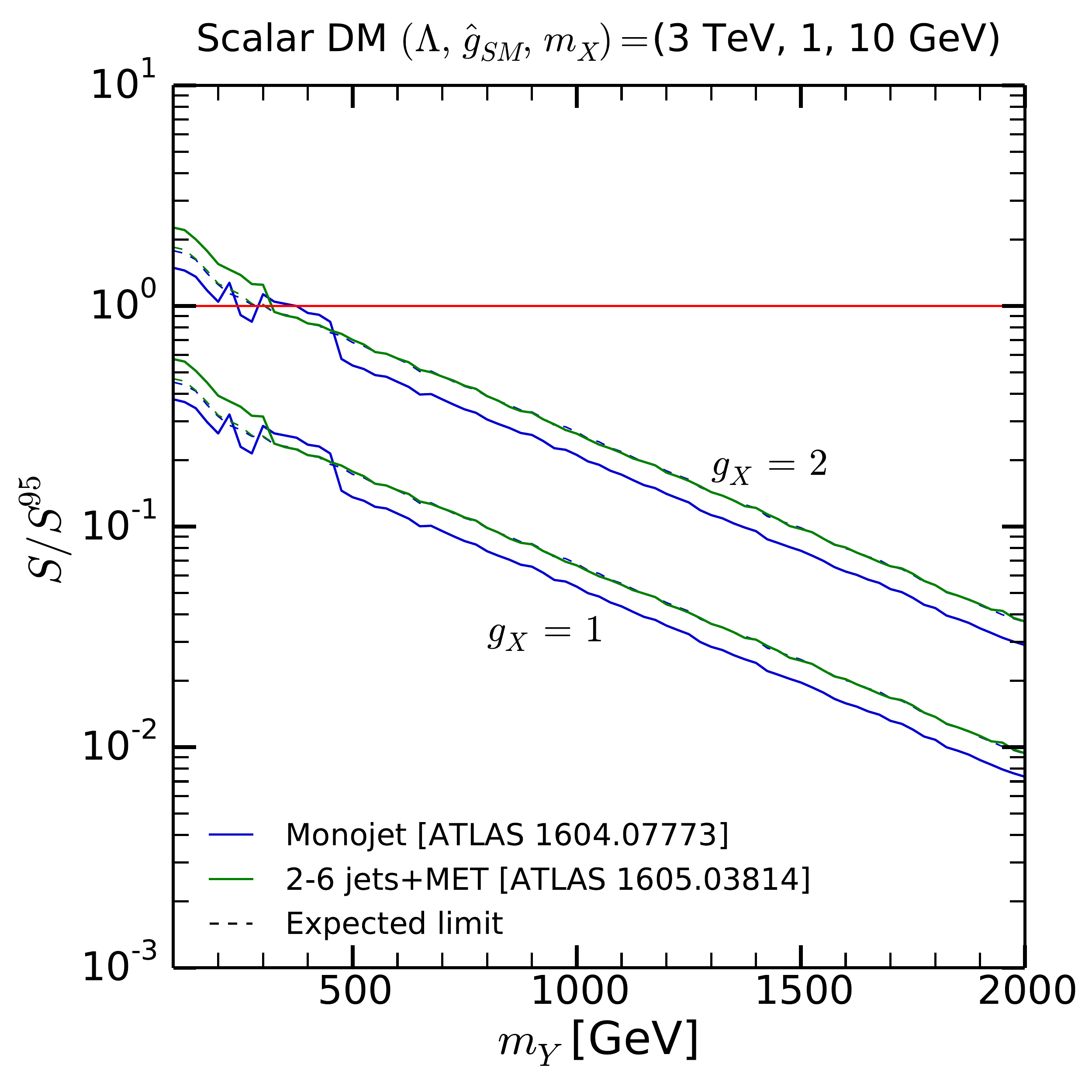}%
 \includegraphics[width=0.33\textwidth]{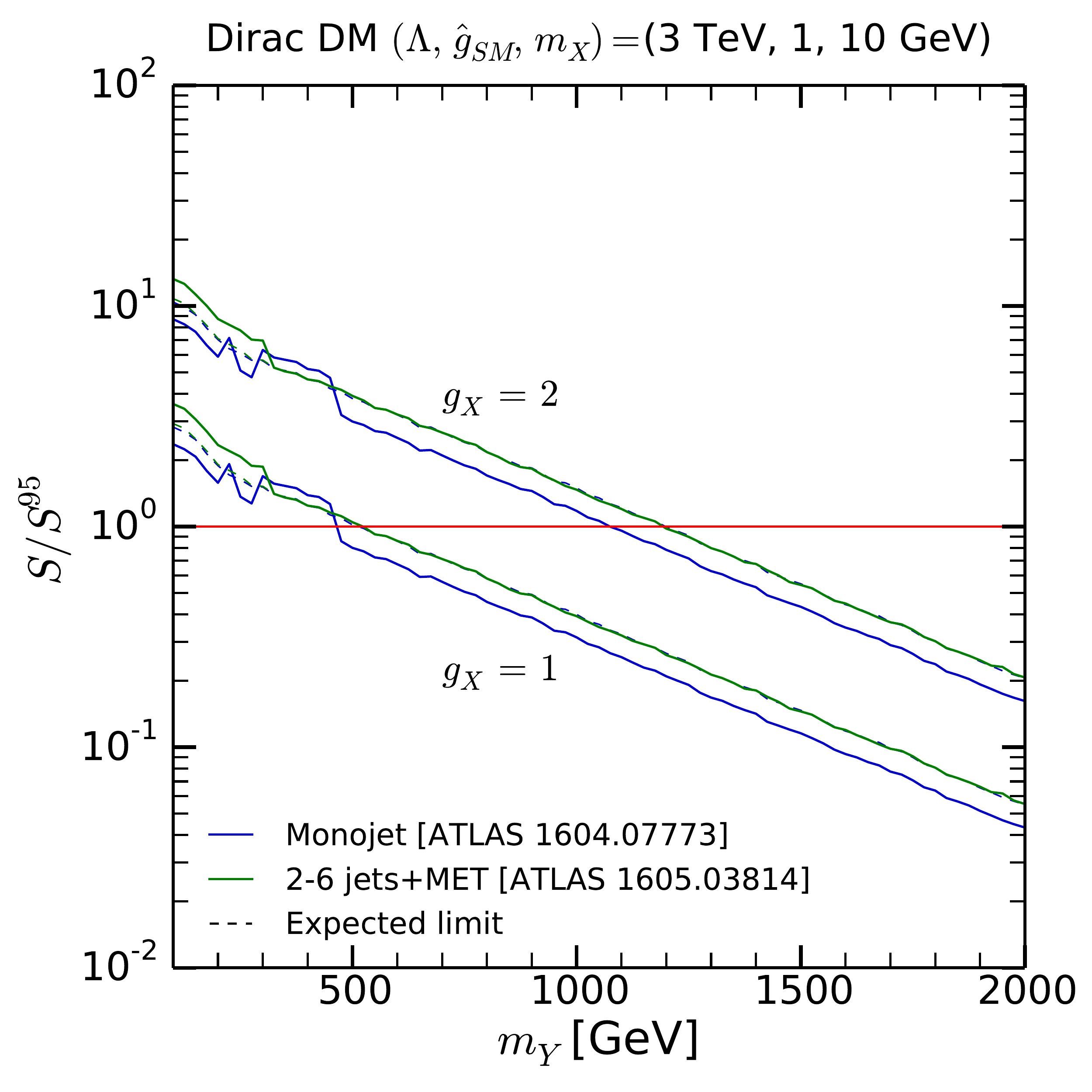}%
 \includegraphics[width=0.33\textwidth]{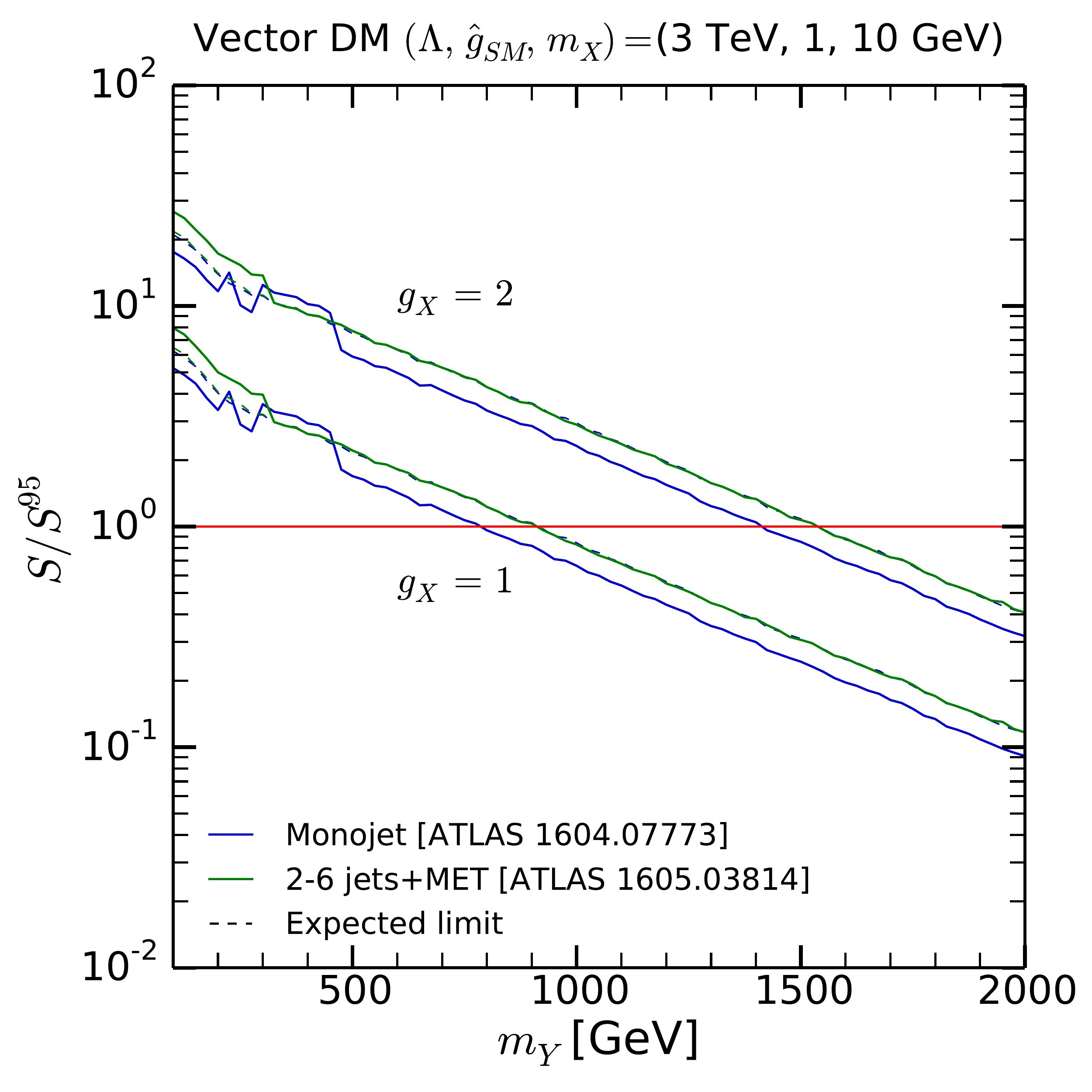}
\caption{Same as Fig.~\ref{fig:exclusion_1d_nu}, but for the leptophobic scenario.}
\label{fig:exclusion_1d}
\end{figure*}

\begin{figure*}\center 
 \includegraphics[width=0.33\textwidth]{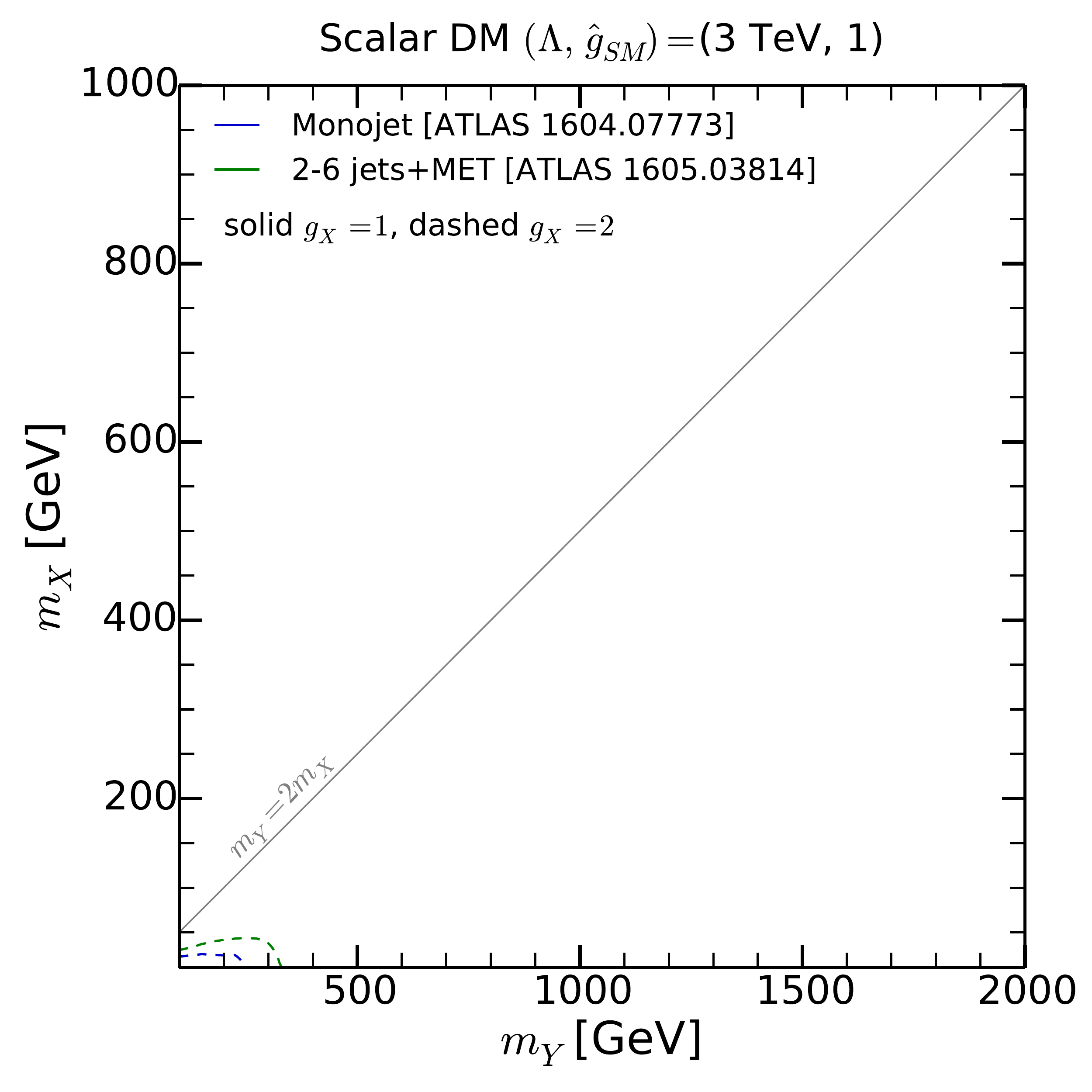}%
 \includegraphics[width=0.33\textwidth]{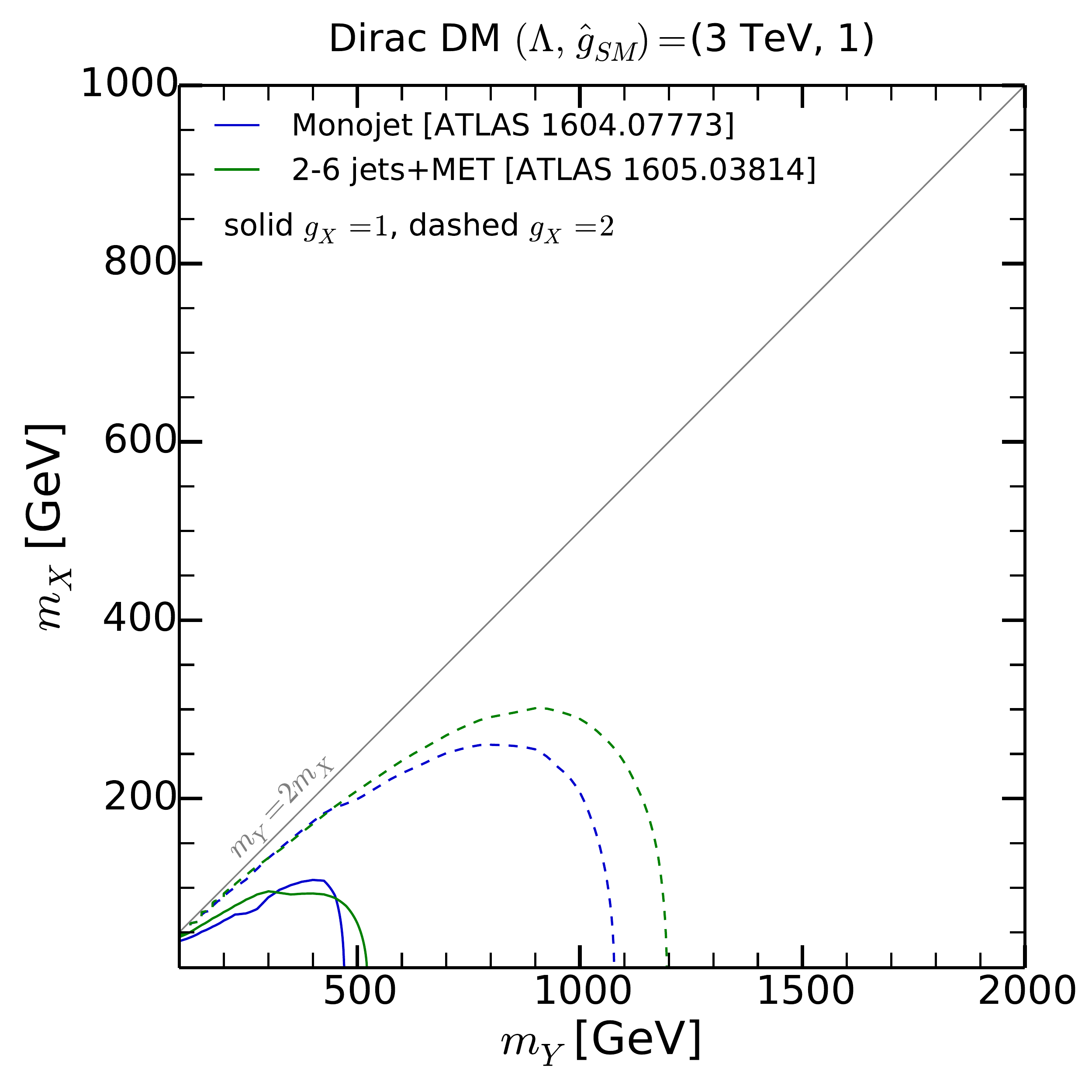}%
 \includegraphics[width=0.33\textwidth]{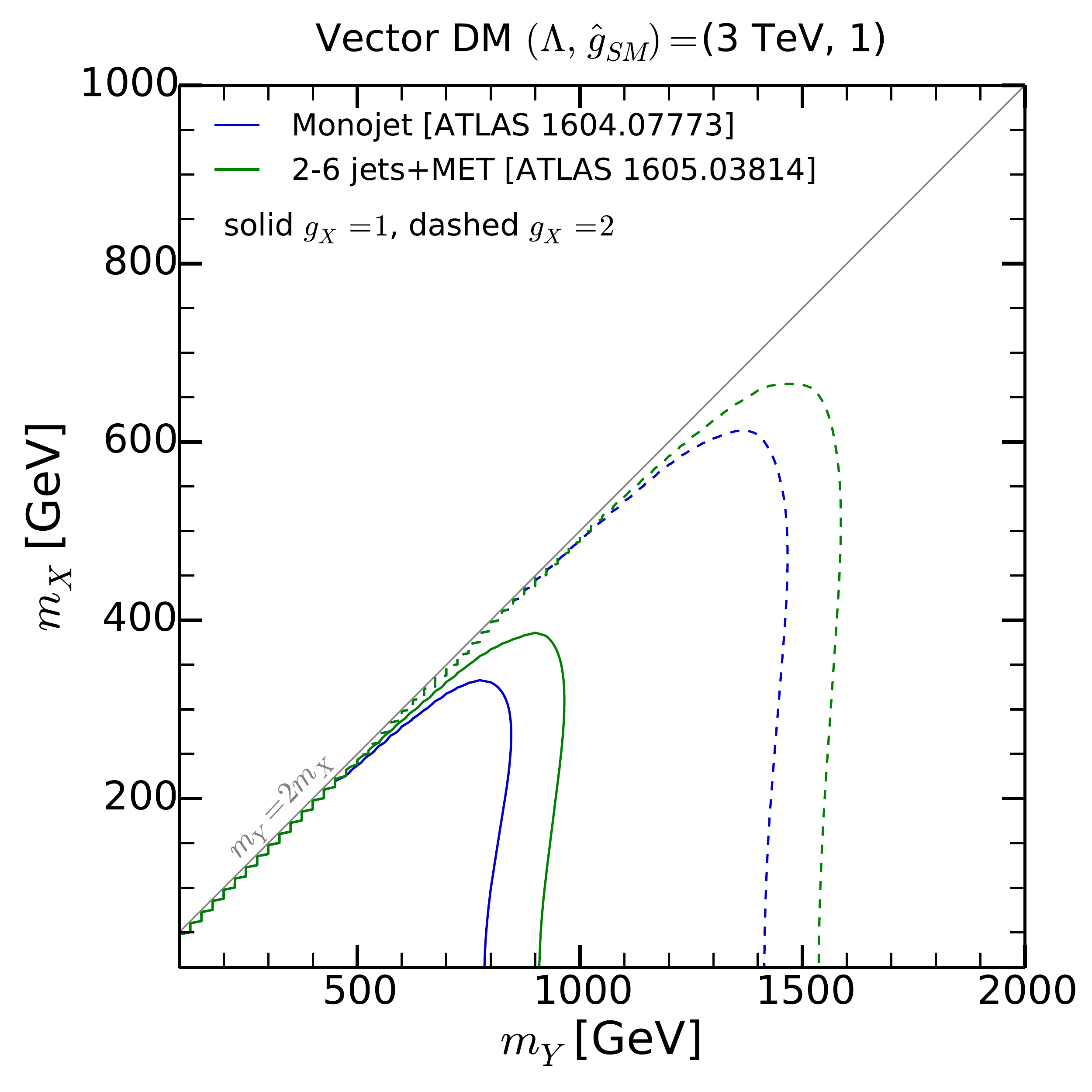}
\caption{Same as Fig.~\ref{fig:exclusion_2d_nu}, but for the leptophobic scenario.}
\label{fig:exclusion_2d}
\end{figure*}

As the production rate scales as $1/\Lambda^2$, the upper limit of $\Lambda$ can be estimated from the plots. 
For instance, for vector DM with $m_Y=100\GeV$, $\Lambda$ should be larger than around 10~TeV for $g_{\rm SM}=g_X=1$.
It should be noted that, due to the $K$ factors of $1.7-1.2$ for $m_Y=100-2000\GeV$ (see Fig.~\ref{fig:xsec-Y2}), these limits are slightly stronger than what would be obtained with LO production rates.

The 95\% CL exclusion in the $m_X$ vs.\ $m_Y$ plane is shown in Fig.~\ref{fig:exclusion_2d_nu}.
Due to the different threshold behaviours, as seen in Eqs.~\eqref{eq:decayS}--\eqref{eq:decayV}, the excluded region near $m_Y=2m_X$ strongly depends on the type of DM. 

We note that we compared the {\sc CheckMATE} results with those obtained by the equivalent analysis implementations in 
{\sc MadAnalysis\,5}~\cite{Conte:2014zja,Dumont:2014tja} (recast codes \cite{ma5:monojet, ma5:multijet})
and {\sc Rivet\,2.5}~\cite{Buckley:2010ar} 
for a couple of representative mass choices and found agreement at the level of 20\% within all three tools. 

The monophoton (as well as mono-$Z/W$) signature could also be interesting to explore the spin-2 model.
However, as seen in Sec.~\ref{sec:production}, the production rate for a pair of DM with a photon is strongly suppressed.  
We checked that there is no constraint for the above benchmark points from the CMS 13~TeV monophoton analysis (12.9~fb$^{-1}$)~\cite{CMS-PAS-EXO-16-039}.

An interesting alternative to the universal coupling $g_{\rm SM}$ is a leptophobic scenario with  
\begin{align}
g^T_\ell\ll \hat g_{\rm SM}\equiv g^T_H=g^T_q=g^T_g=g^T_W=g^T_B\,. 
\end{align}
In this case, the \MET\ signatures come exclusively from decays into DM, because $Y_2$ decays into neutrinos are switched off. Moreover, constraints from dilepton resonance searches, which as we will see in  the next subsection are 
quite severe, are evaded. The results for the leptophobic scenario are presented in Figs.~\ref{fig:exclusion_1d} and \ref{fig:exclusion_2d} in analogy to Figs.~\ref{fig:exclusion_1d_nu} and \ref{fig:exclusion_2d_nu}. 
As expected, the $m_Y<2m_X$ region is no longer constrained. Also, for $g_X=1$, the exclusion becomes considerably weaker for all the DM types; in particular there is no more constraint for scalar DM. 
For $g_X=2$, except scalar DM, the mediator decays into DM dominates the neutrino decay mode even for the universal coupling scenario (see Fig.~\ref{fig:width_dm}), and hence the $m_Y$ limits are very similar.

\subsection{Resonance searches}\label{sec:resonance}

\begin{table*}[t!]\center
\begin{tabular}{l|l|l|l|l|l}
\hline
 decay mode & reference & limit Tab./Fig. & limit on & $\sqrt{s}\,$(TeV) & $L$\,(fb$^{-1}$)  \\ 
\hline\hline
 $jj$ & ATLAS-CONF-2016-069 \cite{ATLAS-CONF-2016-069} & Tab. 2 (Res) 
 & $\sigma({\rm Gaussian})\times B\times A$ & 13 & 15.7 \\
\hline
 $jj(\!+j/\gamma)$ & ATLAS-CONF-2016-070 \cite{ATLAS-CONF-2016-070} & Tab. 4/3 (Res) 
 & $\sigma({\rm Gaussian})\times B\times A$ & 13 & 15.5 \\
\hline
 $WW$ & ATLAS-CONF-2016-062 \cite{ATLAS-CONF-2016-062} & Fig. 6 
 & $\sigma(G_{\rm RS})\times B$ & 13 & 13.2 \\
\hline
 $bb$ & ATLAS-CONF-2016-060 \cite{ATLAS-CONF-2016-060} & Fig. 7(b) (Res) 
 & $\sigma({\rm Gaussian})\times B\times A\times \epsilon_{2b}$ & 13 & 13.3 \\ 
\hline
 $tt$ & CMS-PAS-B2G-15-002 \cite{CMS-PAS-B2G-15-002} & Tab. 4 (1\%) 
 & $\sigma(Z')\times B$ & 13 & 2.6 \\
\hline
 $ZZ$ & ATLAS-CONF-2016-082 \cite{ATLAS-CONF-2016-082} & Fig. 10(d) 
 & $\sigma(G_{\rm RS})\times B$ & 13 & 13.2 \\
\hline
 $\gamma\gamma$ & CMS 1609.02507 \cite{Khachatryan:2016yec} & Fig. 6(middle) 
 & $\sigma(G_{\rm RS})\times B$ & 13+8 & 16.2+19.7 \\
\hline
 $\ell\ell$ & ATLAS-CONF-2016-045 \cite{ATLAS-CONF-2016-045} & Fig. 3(c)  
 & $\sigma(Z')\times B$ & 13 & 13.3 \\ 
\hline
 $hh$ & ATLAS-CONF-2016-049 \cite{ATLAS-CONF-2016-049} & Fig. 11 
 & $\sigma(G_{\rm RS})\times B$ & 13 & 13.3 \\
\hline
\hline  
 $\gamma\gamma$ & ATLAS 1407.6583 \cite{Aad:2014ioa} & Fig. 4, {\sc HepData}~\cite{atlas:diphoton:hepdata}
 & $\sigma(H)\times B\times A$ & 8 & $20.3$ \\
                                & CMS 1506.02301 \cite{Khachatryan:2015qba} & Fig.~6 & $\sigma(G_{\rm RS})\times B$ & 8 & $19.7$ \\
\hline
 $WW$ & ATLAS 1512.05099 \cite{Aad:2015ipg} & Auxiliary Fig. 3  & $\sigma(G_{\rm RS})\times B$ & 8 & $20.3$ \\
\hline
 $ZZ$ & ATLAS 1512.05099 \cite{Aad:2015ipg} & Auxiliary Fig. 4  & $\sigma(G_{\rm RS})\times B$ & 8 &  $20.3$ \\
\hline
\end{tabular}
\caption{Constraints from resonance searches used in this study. 
The observed 95\% CL upper limits on resonant production cross section ($\sigma$) times branching ratio ($B$) (times acceptance ($A$)) from each analysis are 
shown in Fig.~\ref{fig:exclusion_data_resonances} in Appendix~\ref{sec:app2}. 
} 
\label{tab:resonance_search}
\end{table*}

Direct resonance searches can also be used to explore $s$-channel mediator DM models, see \eg~\cite{Chala:2015ama,Arina:2016cqj} for the spin-1 and spin-0 mediator models, respectively. 
Results from Run-II data are already available for a large variety of final states (dijet, dilepton, diphoton, $WW$, $ZZ$, $b\bar b$, $t\bar t$, $hh$) from ATLAS~\cite{ATLAS-CONF-2016-045,ATLAS-CONF-2016-062,ATLAS-CONF-2016-069,ATLAS-CONF-2016-070,ATLAS-CONF-2016-082,ATLAS-CONF-2016-060,ATLAS-CONF-2016-049} and 
CMS~\cite{Khachatryan:2016yec,Sirunyan:2016iap,CMS-PAS-EXO-15-002,CMS-PAS-EXO-16-031,Khachatryan:2016qkc,CMS-PAS-B2G-15-002},  
{\change and give powerful constraints for mediator masses of a few hundred GeV up to several TeV.  
Lower masses are partly covered by Run-I results.}%
\footnote{
{\change We thank the referee for pointing us to the ATLAS analysis \cite{Aad:2014ioa}, which looked for narrow scalar resonances in the diphoton invariant mass spectrum down to 65~GeV.}
}

\begin{figure*}\center 
 \includegraphics[width=0.78\textwidth]{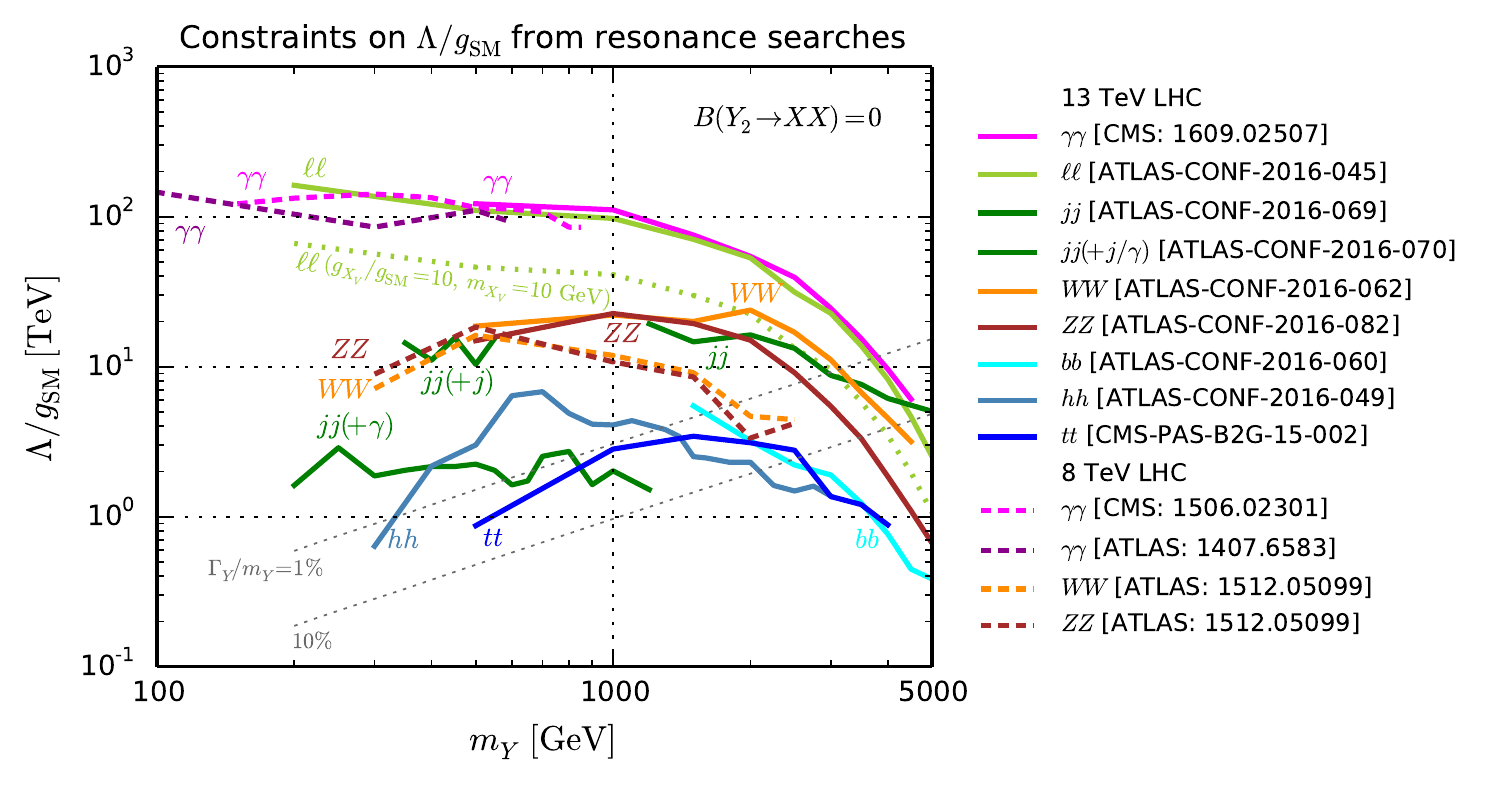} 
\caption{Constraints on $\Lambda/g_{\rm SM}$ from observed 95\% CL upper limits of resonance searches at the {\change 13~TeV (solid) and 8~TeV (dashed)} LHC as a function of the spin-2 mediator mass. 
We assume a negligible branching ratio to DM, except for a dotted line, where 
the vector DM coupling $g_X/g_{\rm SM}=10$ with $m_X=10$~GeV is taken into account as a reference.
Regions below each line are excluded.
Information on the mediator width-to-mass ratio is given by the grey dotted lines. 
}
\label{fig:exclusion_resonances}
\end{figure*}

Table~\ref{tab:resonance_search} lists the {\change current}
resonance search results which we use to constrain our spin-2 simplified model. 
The RS massive graviton is considered in the analyses for pairs of electroweak gauge or Higgs bosons~\cite{Khachatryan:2016yec,ATLAS-CONF-2016-062,ATLAS-CONF-2016-082,ATLAS-CONF-2016-049,Khachatryan:2015qba,Aad:2015ipg} as one of the new physics hypotheses. 
For the fermionic and jet final states in~\cite{ATLAS-CONF-2016-045,ATLAS-CONF-2016-060,CMS-PAS-B2G-15-002,ATLAS-CONF-2016-069,ATLAS-CONF-2016-070}, on the other hand, $Z'$ and a model-independent Gaussian-shaped resonance have been studied.  
Except the dijet and di-$b$-jet analyses {\change at 13~TeV and the low-mass diphoton analysis at 8~TeV from ATLAS}, the limits are provided 
directly on the cross section in the given channel, 
and hence we obtain the model constraints by simply using the $Y_2$ production cross section and the branching ratio discussed in Section~\ref{sec:pheno}.
For the analyses with different hypotheses from the spin-2 resonance, we assume that the acceptance and efficiency are similar.
When limits are given on the fiducial cross section, $\sigma\times B\times A$, we generate LO events normalised by the NLO cross section and apply the fiducial cuts at the parton level by using {\sc MadAnalysis5}~\cite{Conte:2012fm}.

We recall that, for a given mediator mass, the $Y_2$ production cross section depends solely on $g_{\rm SM}/\Lambda$, while the branching ratio depends also on the parameters related to DM, \ie~$g_X$ and $m_X$, as well as on the type of DM.
In the decoupling limit of the dark sector, the constraints on $\Lambda/g_{\rm SM}$ are the most stringent.    
When decays to DM are relevant, the branching ratios to SM particles become smaller and hence the constraints are weakened.  

\begin{figure*}\center 
 \includegraphics[width=0.495\textwidth]{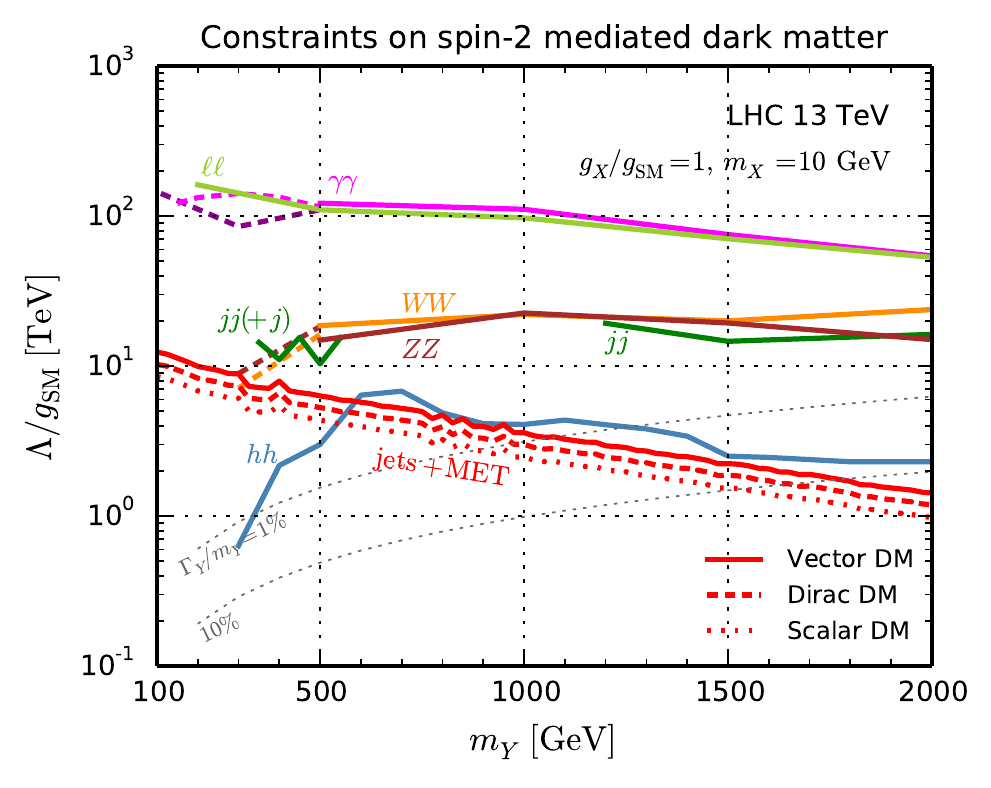}%
 \includegraphics[width=0.495\textwidth]{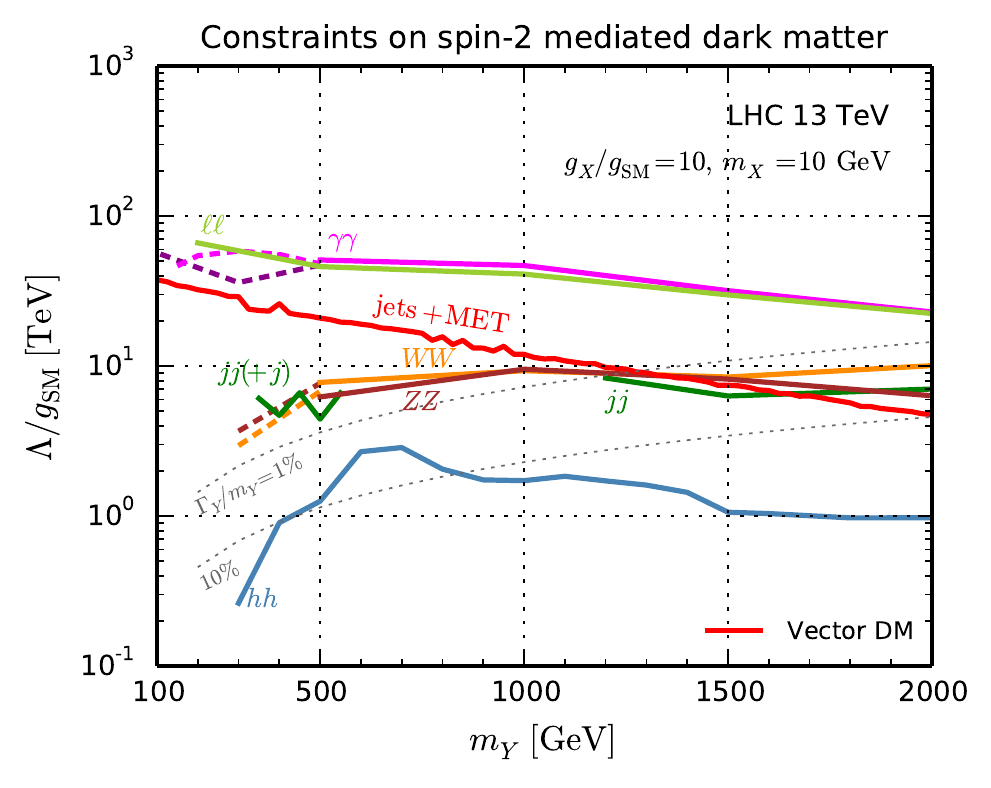} 
\caption{
Summary of the constraints on $\Lambda/g_{\rm SM}$ from searches with and without missing energy at the 13~TeV LHC as a function of the spin-2 mediator mass, for $m_X=10\GeV$ with $g_X/g_{\rm SM}=1$ (left) and 10 (right). 
{\change The labelling of the constraints from resonance searches is the same as in Fig.~\ref{fig:exclusion_resonances}.}
For $g_X/g_{\rm SM}=1$ the differences among the different types of DM for the limits from the resonance searches are not visible.  For $g_X/g_{\rm SM}=10$, however, they are quite relevant so only the vector DM case is shown. 
The figure assumes a universal $g_{\rm SM}$ but is also valid for the leptophobic case when ignoring the $\ell\ell$ lines. }
\label{fig:exclusion}
\end{figure*}

Figure~\ref{fig:exclusion_resonances} shows the constraints on $\Lambda/g_{\rm SM}$ from the observed 95\% CL upper limits of the resonance searches listed in Table~\ref{tab:resonance_search} as a function of the mediator mass, where we assume a negligible branching ratio to DM particles, \ie\ $g_X\ll1$ and/or $m_Y<2m_X$.
Although the branching ratio is small, $B(Y_2\to\gamma\gamma)\sim 4$\% at high mass,   
{\change the diphoton resonance searches 
give the most stringent limit for the whole mass range,
resulting in $\Lambda/g_{\rm SM}\gtrsim 100\TeV$ for $m_Y\lesssim 1\TeV$.
The dilepton channel, also having a branching ratio of about 4\%, provides a similarly strong constraint 
for mediator masses above 200~GeV.}
The dijet and $WW/ZZ$ resonance searches lead to a constraint of a few tens of TeV on $\Lambda/g_{\rm SM}$ for around 1~TeV mediator mass.  
We note again that the limits are obtained based on the NLO production rates which are larger than the LO ones, especially for $pp\to(Y_2\to jj)\gamma$; see Fig.~\ref{fig:xsec-Y2}. 
We also note that, as indicated by grey dotted lines in Fig.~\ref{fig:exclusion_resonances}, 
the mediator width can be very large at high mass and low $\Lambda/g_{\rm SM}$;  
as the experimental analyses often assume a narrow width, this region has to be regarded with caution. 

The weakening of the constraints when $Y_2$ decays into DM are allowed is demonstrated for the dilepton channel in Fig.~\ref{fig:exclusion_resonances}, depicted by a dotted line, where we assume vector DM and take $g_X=10$ and $m_X=10\GeV$.
For instance, at $m_Y=1\TeV$, the dilepton (electron and muon) branching ratio becomes 0.8\%, \ie\ the dilepton production rate becomes smaller by a factor of five, reducing the limit on $\Lambda/g_{\rm SM}$ by $1/\sqrt{5}$.
As seen in Fig.~\ref{fig:width_dm}, the above assumption gives the largest DM branching ratio within the scenarios we consider.%
\footnote{In Fig.~\ref{fig:exclusion_resonances} there is hardly any difference between the $g_X\ll1$ and $g_X=1$ cases.}
{\change Therefore, the diphoton resonance searches, and for $m_Y>200\GeV$ also the dilepton resonance searches, provide stronger constraints on the universal coupling scenario than the searches with missing energy.}

To avoid such severe constraints from resonance searches, it is interesting to consider scenarios beyond the universal coupling case.
{\change The dilepton constraints could be avoided, for example, in the leptophobic scenario, $g^T_\ell=0$, as already discussed in the previous subsection. To avoid the diphoton constraints is somewhat more complicated.
One possibility would be the gravity-mediated DM model~\cite{Lee:2013bua,Lee:2014caa}, where the KK graviton mainly couples to massive particles ---DM, Higgs, massive gauge bosons and top quarks--- while the couplings to photons, gluons and light quarks are highly suppressed. 
In such scenarios, the branching ratios and the production cross sections of the spin-2 resonance strongly depend on the setup and can be very different from those in the universal coupling case. In fact associated production of the mediator with a $W$ or $Z$ boson, or mediator production in vector boson fusion may be more relevant than $s$-channel production in $q\bar q$ or $gg$ fusion.  While such setups can in principle be studied easily in the simplified model framework by appropriately 
choosing the free parameters $g_X^T$ and $g_i^T$ in Eq.~\eqref{smcouplings}, such an analysis is beyond the scope of this paper.} 
A final caveat is that non-universal couplings to gluons and quarks, $g^T_g\ne g^T_q$, give rise to a unitarity violating behaviour at higher order in QCD~\cite{Artoisenet:2013puc}.
We therefore only consider phenomenological scenarios with $g^T_g=g^T_q$.

\section{Summary}\label{sec:summary}

We considered a simplified DM model where the DM candidate couples to the SM particles via an $s$-channel spin-2 mediator, $Y_2$, and studied the constraints from the current 
LHC data. In particular, we compared the constraints from searches with and without missing energy. 

For universal couplings of the mediator to SM particles, we found that 
{\change diphoton resonance searches provide the strongest constraints, $\Lambda/g_{\rm SM} \gtrsim 100$ TeV for $Y_2$ masses up to $\sim 1$~TeV.  
For $\Lambda/g_{\rm SM}=10$ (3)~TeV, the exclusion extends up to 4 (beyond 5) TeV in $m_Y$. 
The dilepton channel provides a similarly strong constraint for mediator masses above 200 GeV. 
Monojet and multijet+\MET\ searches are competitive only if the mediator decays into photons and leptons 
are heavily suppressed; in this case they could provide complementary constraints to the other resonance 
searches in particular in the low-mass region below 0.5--1~TeV, depending on $g_X/g_{\rm SM}$. }

For $m_Y<2m_X$, \MET\ signatures arise solely from $Y_2$ decays into neutrinos, leading to $m_Y\gtrsim 700$~GeV 
{\change for $g_X/\Lambda=g_{\rm SM}/\Lambda=(3\TeV)^{-1}$, based on 3.2~fb$^{-1}$ of data at $\sqrt{s}=13$~TeV.}
For $m_Y>2m_X$, the limit crucially depends on $g_X$ and the type of dark matter. The dependence on the DM mass is less pronounced unless one approches the threshold region. For $m_X=10$~GeV and $g_X/\Lambda=g_{\rm SM}/\Lambda=(3\TeV)^{-1}$, we found $m_Y\gtrsim 750$, 950, and 1100~GeV for scalar, Dirac, and vector DM, respectively. This increases to 850, 1300, and 1550~GeV when doubling $g_X$.
We note that the obtained limits are based on the NLO-QCD predictions, which give a larger production rate than at the LO. 
The $K$ factor depends on the mediator mass and the production channel.

The complementarity among the different searches is illustrated in Fig.~\ref{fig:exclusion}, where we have rescaled the reach of the jets + \MET\  searches from $3.2$ to $15$~fb$^{-1}$ in order to make a fair comparison. We see that, for the same amount of data,  
in case of $g_X\simeq g_{\rm SM}$ the missing energy searches are roughly competitive with the dijet and heavy diboson ($WW$, $ZZ$) searches, pushing $\Lambda/g_{\rm SM}$ beyond 20~TeV. (As mentioned, when the dilepton and diphoton constraints hold, they give even stronger limits.) 

For $g_X/g_{\rm SM}=10$ (or $g_X/\hat g_{\rm SM}=10$), also the resonance constraints strongly depend on the type of DM. Therefore, in the right plot in Fig.~\ref{fig:exclusion} only the vector DM case is shown. We see that the jets+\MET\ searches give stronger constraints than the dijet and heavy diboson searches up to mediator masses of about $1.2$~TeV. The dilepton and diphoton constraints 
are weakened by about a factor of two but still give the strongest constraints. 

We hope our work will be useful to find reasonable benchmark scenarios for spin-2 mediated DM searches at the LHC as well as to construct viable UV-completed models which can give predictions for those parameters. 
We also note that our study on resonance searches in Sec.~\ref{sec:resonance} can be applied not only for spin-2 mediated DM models but also for usual RS-type graviton searches; see also, \eg~\cite{Alvarez:2016ljl}.
{\change As a final remark we like to point out that in a full model the presence of KK excitations might alter the LHC phenomenology as compared to the simplified model scenarios discussed here. Examples are limits on gauge KK modes 
providing additional constraints on light gravitons, or KK excitations of the DM fields contributing to \MET\ signatures. 
While this goes well beyond the simplified model picture, it is certainly an interesting topic for future studies.}

\section*{Acknowledgements}

We would like to thank 
G.~Das, C.~Degrande, V.~Hirschi and H-S.~Shao for help with the NLO calculations, 
and M-H.~Genest, F.~Maltoni, V.~Sanz and M.~Zaro for valuable discussions.
We are also thankful to C.~Doglioni and K.~Krizka for discussions on ATLAS-CONF-2016-070.

This work was supported in part by the French ANR, project DMAstro-LHC ANR-12-BS05-0006. 
U.\,L.\ is supported by the {\it Investissements d'avenir}, Labex ENIGMASS. 
K.\,M.\ is supported by the Theory-LHC-France Initiative of the CNRS (INP/IN2P3). 
K.\,Y.\ acknowledges support for a long-term stay at LPSC Grenoble from the Program for Leading Graduate Schools of Ochanomizu University; she also thanks the LPSC Grenoble for hospitality while this work was completed.

\appendix
\section{Supplemental material for recasting}\label{sec:app}

\subsection{Searches with missing energy}
\label{sec:app1}

\begin{figure}\center
 \includegraphics[width=1.\columnwidth]{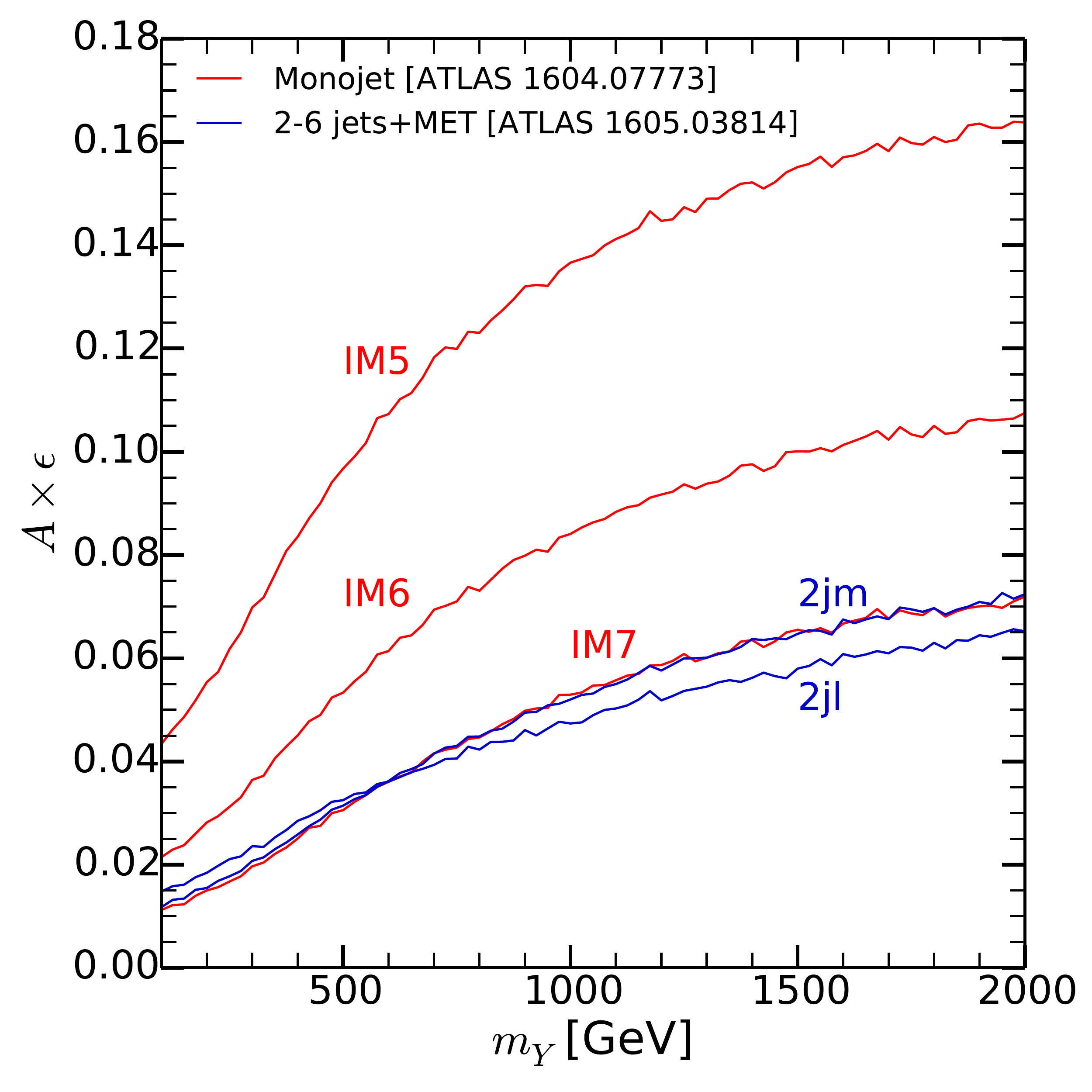}
\caption{Signal acceptance times efficiency, $A\times \epsilon$, as a function of the mediator mass for the most relevant SRs, \ie\ IM5, IM6 and IM7 of the ATLAS monojet search~\cite{Aaboud:2016tnv} and 2jl and 2jm of the ATLAS 2--6 jets + \MET\ search~\cite{Aaboud:2016zdn}, evaluated with {\sc CheckMATE2}~\cite{Dercks:2016npn}.}
\label{fig:efficiencies}
\end{figure}

As mentioned in the main part of the paper, in case of the monojet and the 2--6 jets + \MET\ searches, the signal comes solely from on-shell mediator production with the $Y_2$ decaying into neutrinos and/or DM. The signal selection efficiency (more precisely acceptance times efficiency, $A\times \epsilon$)  depends only on the properties of the mediator, but not on those of the invisible decay products. 
Figure~\ref{fig:efficiencies} shows $A\times \epsilon$ 
for those SRs which, depending on $m_Y$, can be the most sensitive ones in each of the two ATLAS analyses considered in this paper.   
As a service to the reader and potential user of our work, the complete $A\times \epsilon$ tables for all SRs are available in numerical form at \cite{recasting:lpsc}. 

\subsection{Resonance searches}
\label{sec:app2}

\begin{figure*}\center 
 \qquad
 \includegraphics[height=0.44\textwidth]{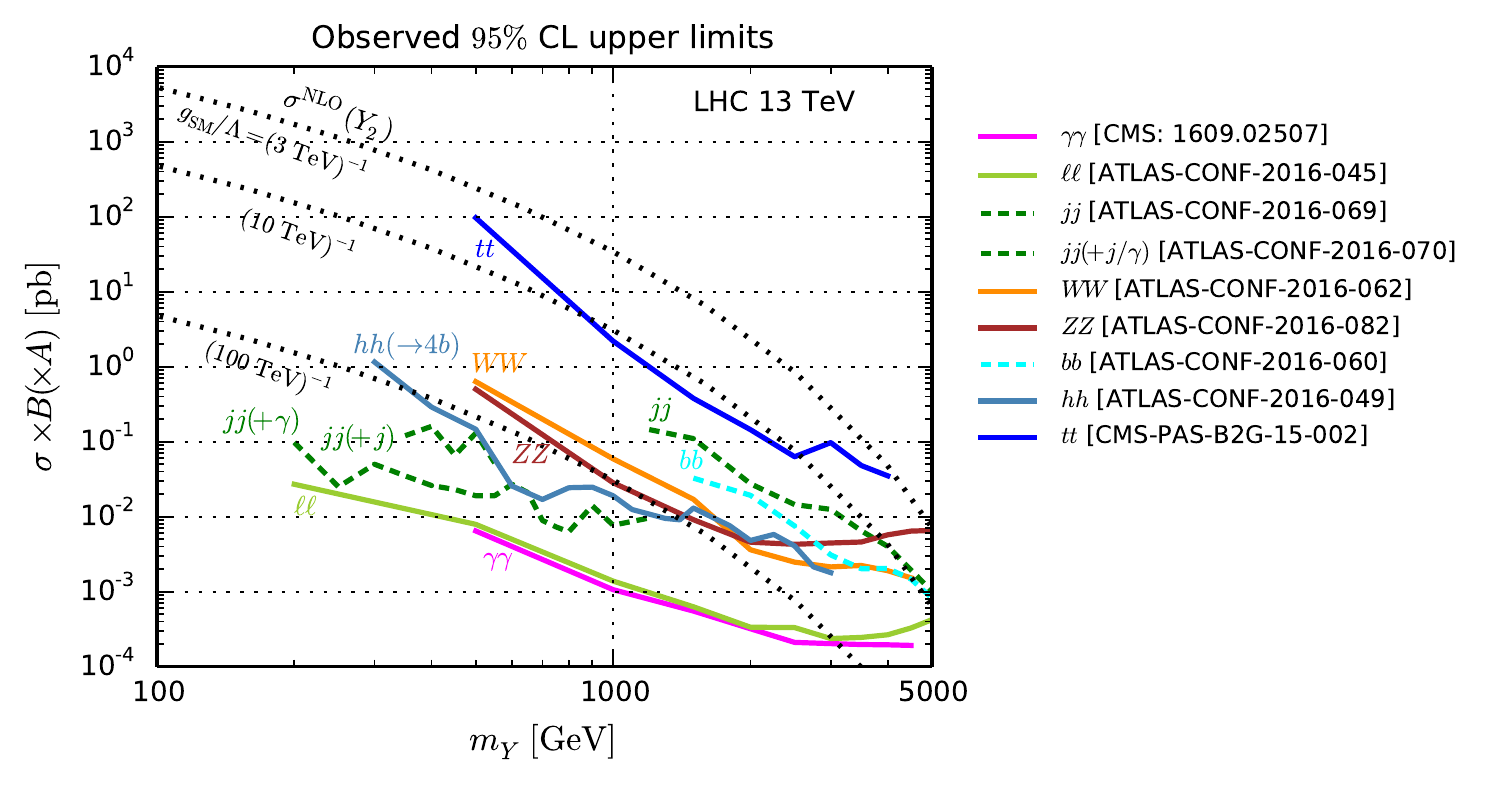} 
 \includegraphics[height=0.44\textwidth]{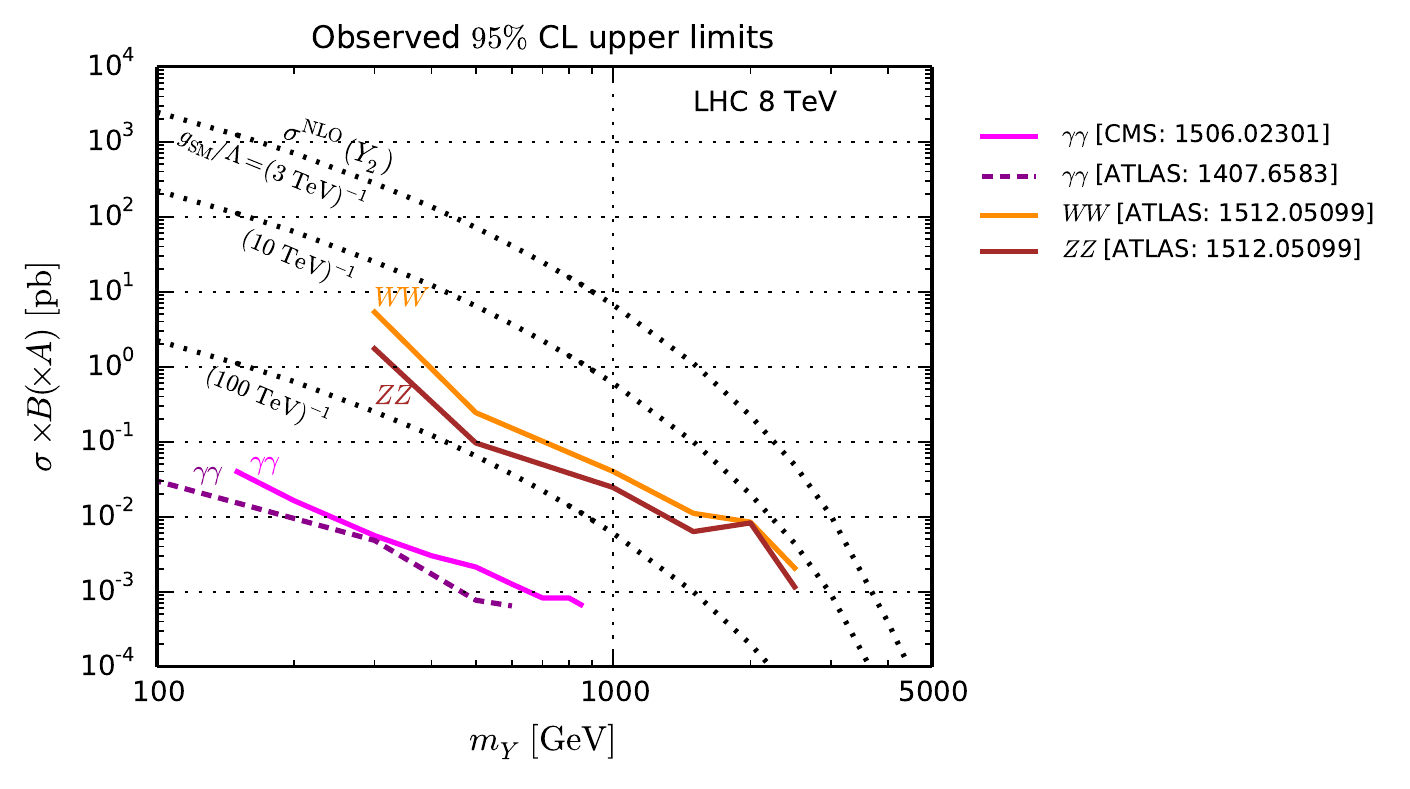} 
\caption{Observed 95\% CL upper limits on resonant production cross section times branching ratio (times acceptance) as a function of the resonance mass from each experimental paper; see Table~\ref{tab:resonance_search} for more detailed information.
Dashed lines denote limits including cut acceptance. For reference, NLO production cross sections of the spin-2 mediator are shown by dotted lines for different values 
of $g_{\rm SM}/\Lambda$.}
\label{fig:exclusion_data_resonances}
\end{figure*}

In Fig.~\ref{fig:exclusion_data_resonances} we show observed 95\% CL upper limits on resonant production cross section times branching ratio (times acceptance) as a function of the resonance mass from each experimental paper.
The analyses denoted by solid lines present the limit on $\sigma\times B$, while those by dashed lines provide the limit on $\sigma\times B\times A\,(\times \epsilon\ {\rm for}\ b\bar b)$; see Table~\ref{tab:resonance_search} for more detailed information.

As indicated in Table~\ref{tab:resonance_search}, the dijet (+ ISR jet/photon) and $t\bar t$ analyses at 13~TeV as well as the   
ATLAS 8~TeV diphoton analysis provide tables with the numbers corresponding to the lines in the exclusion plots, which is very convenient for our purpose. 
The other analyses do not provide explicit values, and hence we have to extract these data from the exclusion plots `by hand', \eg\ using {\sc WebPlotDigitizer}~\cite{wepplotdigitizer}, a public software.
To avoid that other people have to redo this exercise, our digitised data files are available at \cite{recasting:lpsc}  
and on the new {\sc{PhenoData}} database~\cite{phenodata}. 
We encourage the experimental collaborations to provide digitised data together with their plots, in order to make it easier to use their results.  

Finally, we notice a caveat regarding the re-interpretation of the low-mass resonance search in dijet plus ISR final states~\cite{ATLAS-CONF-2016-070}. 
We found that final state radiation (FSR) may be also important and give rise to a non-trivial structure in the dijet invariant mass spectrum. Technically, simulated event shapes can differ by including FSR or not in the matrix elements, which may affect the parameter fitting procedure for a bump search.

\bibliographystyle{JHEP}
\bibliography{bibdm}

\end{document}